\documentclass[twocolumn, times, tighten, twocolappendix, astrosymb]{aastex63}
\submitjournal{ApJ}
\accepted{2025 April 4}
\usepackage{hyperref}
\usepackage{amsmath}
\usepackage{amssymb}
\usepackage{graphicx}
\usepackage{color}
\usepackage{longtable}
\usepackage{lineno}
\usepackage[utf8]{inputenc}
\DeclareUnicodeCharacter{2212}{-}

\newcommand{\kms}{kms$^{-1}$}

\newcommand{\av}{$A_{\rm V}$}
\newcommand{\logg}{$\log(g)$}
\newcommand{\hst}{\textit{HST}}
\newcommand{\jwst}{\textit{JWST}}

\newcommand{\zsun}{$Z_{\odot}$}
\newcommand{\msun}{$M_{\odot}$}
\newcommand{\rsun}{$R_{\odot}$}

\newcommand{\disappear}[1]{}

\newcommand{\corr}[1]{\color{black}#1}

\begin{document}

\title{A Low Metallicity Massive Contact Binary Star System Candidate in WLM identified by Hubble and James Webb Space Telescope imaging}

\author[0000-0003-3747-1394]{Maude Gull}
\affiliation{Department of Astronomy, University of California, Berkeley, Berkeley, CA 94720, USA}

\author[0000-0002-6442-6030]{Daniel R. Weisz}
\affiliation{Department of Astronomy, University of California, Berkeley, Berkeley, CA 94720, USA}

\author[0000-0002-6871-1752]{Kareem El-Badry}
\affiliation{Department of Astronomy, California Institute of Technology, Pasadena, CA 91125, USA}

\author[0000-0003-4307-0060]{Jan Henneco}
\affiliation{Heidelberger Institut für Theoretische Studien, Schloss-Wolfsbrunnenweg 35, 69118, Heidelberg, Germany}
\affiliation{Universität Heidelberg, Im Neuenheimer Feld 226, 69120, Heidelberg, Germany}

\author[0000-0002-1445-4877]{Alessandro Savino}
\affiliation{Department of Astronomy, University of California, Berkeley, Berkeley, CA 94720, USA}

\author[0000-0001-7531-9815]{Meredith Durbin}
\affiliation{Department of Astronomy, University of California, Berkeley, Berkeley, CA 94720, USA}

\author[0000-0003-1680-1884]{Yumi Choi}
\affiliation{NSF National Optical-Infrared Astronomy Research Laboratory, 950 North Cherry Avenue, Tucson, AZ 85719, USA}

\author[0000-0002-2970-7435]{Roger E. Cohen}
\affiliation{Department of Physics and Astronomy, Rutgers the State University of New Jersey, 136 Frelinghuysen Rd., Piscataway, NJ, 08854, USA}

\author[0000-0003-0303-3855]{Andrew A. Cole}
\affiliation{Greenhill Observatory \& School of Natural Sciences, University of Tasmania, Private Bag 37 Hobart, Tasmania 7001 Australia}

\author[0000-0001-6464-3257]{Matteo Correnti}
\affiliation{INAF Osservatorio Astronomico di Roma, Via Frascati 33, 00078, Monteporzio Catone, Rome, Italy}
\affiliation{ASI-Space Science Data Center, Via del Politecnico, I-00133, Rome, Italy}

\author[0000-0002-1264-2006]{Julianne J.~Dalcanton}
\affiliation{Center for Computational Astrophysics,
Flatiron Institute,
162 Fifth Ave,
New York, NY, 10010, USA}
\affiliation{Department of Astronomy,
University of Washington,
Box 351580,
Seattle, WA, 98195, USA}

\author[0000-0003-0394-8377]{Karoline M. Gilbert}
\affiliation{Space Telescope Science Institute, 3700 San Martin Dr., Baltimore, MD 21218, USA}
\affiliation{The William H. Miller III Department of Physics \& Astronomy, Bloomberg Center for Physics and Astronomy, Johns Hopkins University, 3400 N. Charles Street, Baltimore, MD 21218, USA}

\author[0000-0002-8937-3844]{Steven R. Goldman}
\affil{Space Telescope Science Institute, 3700 San Martin Drive, Baltimore, MD 21218, USA}

\author[0000-0001-8867-4234]{Puragra Guhathakurta}
\affiliation{Department of Astronomy \& Astrophysics, University of California Santa Cruz, 1156 High Street, Santa Cruz, CA 95064, USA}

\author[0000-0001-5538-2614]{Kristen B.~W. McQuinn}
\affiliation{Space Telescope Science Institute, 3700 San Martin Drive, Baltimore, MD 21218, USA}
\affiliation{Department of Physics and Astronomy, Rutgers, The State University of New Jersey, 136 Frelinghuysen Rd, Piscataway, NJ 08854, USA}

\author[0000-0002-8092-2077]{Max J. B. Newman}
\affiliation{Department of Physics and Astronomy, Rutgers, The State University of New Jersey, 136 Frelinghuysen Rd, Piscataway, NJ 08854, USA}

\author[0000-0003-0605-8732]{Evan D. Skillman}
\affiliation{Minnesota Institute for Astrophysics, University of Minnesota, 116 Church Street South East, Minneapolis, MN 55455, USA}

\author[0000-0002-7502-0597]{Benjamin F. Williams}
\affiliation{Department of Astronomy, University of Washington, Box 351580, U.W., Seattle, WA 98195-1580, USA}

\begin{abstract}

We present archival \hst\ and \jwst\ ultraviolet through near infrared time series photometric observations of a massive minimal-contact binary candidate in the metal-poor galaxy WLM ($Z = 0.14 Z_{\odot}$). This discovery marks the lowest metallicity contact binary candidate observed to date. We determine the nature of the two stars in the binary by using the eclipsing binary modeling software (PHysics Of Eclipsing BinariEs; \texttt{PHOEBE}) to train a neural network to fit our observed panchromatic multi-epoch photometry. The best fit model consists of two hot MS stars ($T_1=29800^{+2300}_{-1700}$~K, $M_1=16^{+2}_{-3}$~\msun, and $T_2=18000^{+5000}_{-5000}$~K, $M_2=7^{+5}_{-3}$~\msun). We discuss plausible evolutionary paths for the system, and suggest the system is likely to be currently in a contact phase. Future spectroscopy will help to further narrow down evolutionary pathways. This work showcases a novel use of data of \jwst\ and \hst\ imaging originally taken to characterize RR Lyrae. We expect time series imaging from LSST, BlackGEM, etc. to uncover similar types of objects in nearby galaxies. 

\end{abstract}


\section{Introduction}

Massive metal-poor binary star systems ($M_{\star}>8$\msun, $Z_{\star}< 0.2\ Z_{\odot}$) are key to many open problems in astrophysics. At high-redshift, these massive binary systems may contribute significantly to the ultra-violet (UV) and optical spectral energy distributions (SEDs) of star-forming galaxies \citep[e.g.,][]{steidel16,Eldridge20,eldridge22}. Across all redshifts, they are thought to be the likely progenitors for many gravitational wave sources \corr{\citep[e.g.,][and references therein]{Spera15,Abbott16,Briel23,Marchant23,Dorozsmai24}.} 

Studies over the last decade have shown that most massive stars are in binary systems and will interact with their companions during their evolution \citep[e.g.,][]{Sana2013,Rizzuto13,Kobulnicky14,moe15,Dunstall15,Moe17,Mahy22,Offner23,Marchant23}. A star's evolutionary path is dramatically altered by the presence of a close companion \citep[e.g.,][]{Pacznski67, Podsiadlowski92,Langer12,Sana2013,Dunstall15,Willcox23}. Interacting massive stars can lead to a variety of astrophysical phenomena such as X-ray binaries \citep[e.g.,][]{Jones73,verbunt93,Tauris06}, stripped-envelope supernovae \corr{\citep[][]{wellstein99,Eldridge13,Tauris15,Laplace21}}, and gravitational waves sources \citep[e.g.,][]{schneider01,Tauris17,Mandel22}, and they can play an important role in nucleosynthesis \citep[e.g.,][]{Lattimer76,Kasen17,Margutti21}.

While there has been a tremendous effort to study massive binaries in the Local Group (LG), the lowest metallicity environment that has received significant attention is the SMC \citep[e.g,][]{Moe13,Hilditch05,lamb16,Mahy20,Dufton19,Bodensteiner21,Rickard23,Shenar24}. 
The physics of massive binary star systems (e.g., rotation, mass-loss) and the formation and evolution of their descendants (e.g., supernovae, gravitational wave sources) are likely to depend on metallicity such that the SMC is not sufficiently metal-poor to characterize the full extent of metallicity effects \citep[e.g.,][]{Klencki18,Gull22,Ramachandran24,telford24,Shenar24}. Despite the clear importance of metallicity, there are very few observational studies for massive binaries in the sub-SMC metallicity regime \citep[$Z_{\star} \leq 0.2$~\zsun;][]{Bodensteiner21}. Observing resolved massive stars in sub-SMC metallicity environments is challenging, due to large typical distances to star-forming dwarf galaxies ($D\gtrsim1$~Mpc) with such metallicities, hence why only a few quantitative studies exist \citep[e.g.,][]{telford21,Gull22,telford23}. 
 Acquiring sufficiently high SNR spectra and photometry, even with our most powerful space- and ground-based facilities, is challenging owing to faintness and crowding in these environments. These issues are compounded for the study of time variations in spectra or photometry (i.e., time series data for eclipsing systems) for which high SNR data is needed for each epoch.

In this paper, we use the combined power of \hst\ and \jwst\ to provide the first detailed time series studies of a massive binary system in a sub-SMC metallicity dwarf galaxy. Specifically, it is the first study of a massive contact binary candidate in a galaxy more distant and metal-poor than the SMC. A contact binary is a pair of stars that are sufficiently close together such that they physically touch one another, i.e., both overfill their Roche Lobes, which are the Roche equipotential surfaces through the first Lagrange point (L1; \citep[e.g.,][]{Pols94,Wellstein01,Menon21}. 

Massive contact binaries are complex systems, and their evolution, especially after the onset of the contact phase, is being actively explored, both theoretically \citep[e.g.,][]{demink07, Marchant16,Hastings20,Fabry22,Fabry23,Henneco2024} and observationally \citep{Penny08,lorenzo14,almeida15,Martins17,Rauc17,Abdul-Masih19}. While several candidates have been identified in surveys of the LMC and SMC \citep[e.g.,][]{Balona92,Muraveva14,Pawlak16}, only $\sim 10$ contact binaries have been studied in detail in the LMC ($\sim 50 \% Z_{\odot}$) and $\sim 12$ in the SMC ($\sim 20 \% Z_{\odot}$) \citep{Hilditch05,North10,Mahy20,AbdulMasih21,Rickard23,Wu23,Wu23b}. Owing to the observational challenges described above, none have been identified in more distant and/or metal-poor galaxies. 

We focus on the galaxy Wolf–Lundmark–Melotte (WLM; also known as DDO221). WLM is a nearby, isolated dwarf with active star-formation \citep[$D\sim 1$~Mpc, log(SFR) $\sim -2.24$~\msun\ yr$^{−1}$;][]{Karachentsev13} with an average gas-phase metallicity of $\sim 14 \% \, Z_{\odot}$ \citep[$12+log(O/H) = 7.83$;][]{Lee05}. Several studies have explored aspects of its rich population of metal-poor massive stars \citep{massey07,Urbaneja08,tramper11,Tramper14,Goldman19,Gonzalestora21,telford21,Bonanos23}, but, aside from Cepheids as distance indicators \citep{Lee21}, none have explicitly focused on time series analysis of its massive star content.

Here, we have identified candidate massive contact binary systems (WLM-CB1) in time series imaging from archival \hst\ and \jwst\ observations that were originally designed for the detection of RR Lyrae stars. Our goals in the following sections are to provide the first observational data point of massive contact binaries in a metal-poor galaxy beyond the Magellanic Clouds and to demonstrate the type of data and analysis needed to identify and characterize these types of systems in other nearby galaxies. 

This paper is organized as follows. \S \ref{sec:data} presents our archival \hst\ and \jwst\ Early Release Science observations of WLM-CB1. \S \ref{sec:lc} describes our analysis of the multi-band light curve observation, and using MESA. Finally, we discuss the results of our analysis and the future observations needed to confirm the nature of WLM-CB1 in \S \ref{s:discussion}. 

Throughout this analysis we adopt the tip of the red giant branch distance measurement of 968$^{+5}_{-7}$~kpc ($\mu=24.93$~mag) from \citet{Albers19}.


\begin{figure}[th!]
\includegraphics[width=\columnwidth]{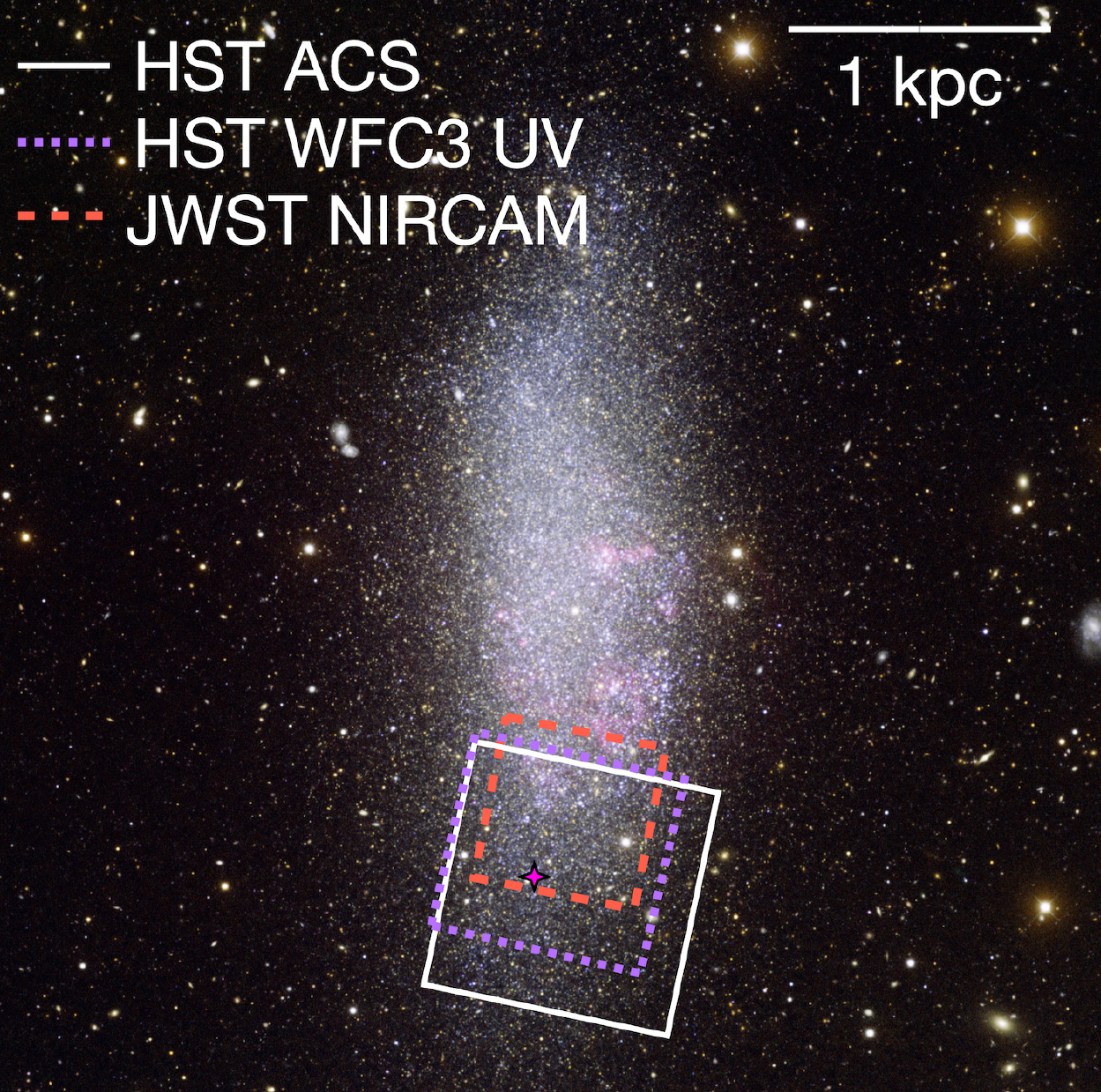} 
 \caption{An optical ground-based image of WLM (credit: NOIRLab/NSF/AURA/CTIO/Local Group Survey Team and T.A. Rector (University of Alaska Anchorage)) with the \hst/ACS, \hst/WFC3, and \jwst/NIRCAM footprints overplotted. 
The location of WLM-CB1 (RA: 00 02 00.23, DEC:$−$15 31 05.20) is shown as a pink star. It falls within all three footprints.}\label{pos}
\end{figure}

\section{Photometry} \label{sec:data}

To investigate the system (WLM$-$CB1), we use a combination of archival \hst\ data and \jwst\ early release science imaging. The footprints for our datasets are shown in Figure~\ref{pos}.

\subsection{\hst\ Imaging and Photometry}

\begin{figure*}[ht!]
\centering
\includegraphics[width=16.5cm]{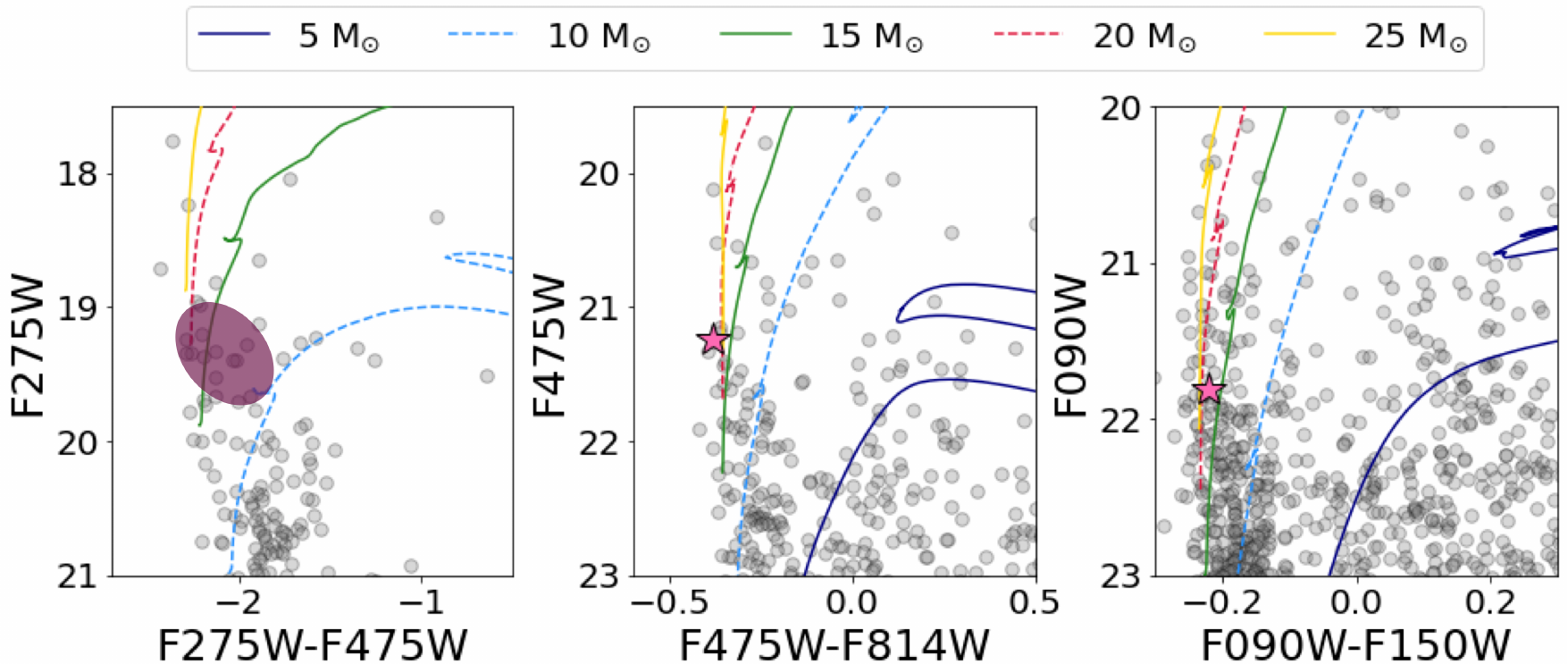} 
 \caption{Select \hst\ and \jwst\ CMDs of WLM, zoomed in the upper MS: UV-optical (left), optical-only (middle), near-IR (right). WLM-CB1 is shown as a star symbol. The optical and near-IR values of WLM-CB1 reflect the maximum point on the optical and near-IR light curves i.e., maximum brightness of the system. Since we have no direct constraint initially on the the phase of the UV data point, we indicate likely CMD position by the shaded oval. We overplot select MIST tracks ([Fe/H]=$-0.82$, v/vcrit = 0.4, $A_v$ = 0.10) for reference. Note that the 5 \msun\ track falls off in the UV-optical CMD, due to its faintness in F275W. A visual comparison with the single-star models suggests a lower limit of $\sim10$~ \msun\ on the current mass of WLM-CB1. 
} \label{cmd}
\end{figure*}

We use archival \hst\ ACS/WFC imaging of WLM that was originally acquired for deep photometry and RR Lyrae studies (GO$-$13768; PI Weisz; \citealt{Albers19}). These multi-epoch observations were carried out over a period of roughly 3 days. There are 24 exposures with ACS/WFC F475W and 26 exposures with ACS/WFC F814W. We summarize the observations in Table \ref{timeseries}.

We further complement our dataset with multi-band photometry observed as part of a larger survey that includes LG dwarf galaxies (GO$-$15275; PI Gilbert). 
The WLM multi-band photometry additionally covers the following bands: WFC3/UVIS F275W and F336W which are part of the Local Ultraviolet to Infrared Treasury (LUVIT, \citep[LUVIT][]{Gilbert25}) dataset. We use individual exposures of the WFC3/UVIS F275W and F336W photometric observations to obtain short NUV time series photometry from the dataset. Observations in each filter cover a time baseline of $\sim$30 min, are summarized in Table \ref{timeseries} as well.
 
The ACS/WFC F475W and ACS/WFC F814W multi-epoch observations were reduced using DOLPHOT \citep{Dolphin00,DOLPHOT}. For a given reference image footprint, DOLPHOT takes any exposure and reports a measurement of point-spread function photometry for every detected object. This method has been well-tested by previous LG \hst\ resolved star studies \citep[e.g.,][]{Dalcanton09,Williams14,Williams19,Savino22,Gull22,Savino23}{}. We follow the same steps as \citet{Savino22}, who base their DOLPHOT setup on \citet{Williams14}. For a more detailed discussion of the reduction we refer the reader to \citet{Williams14} and \citet{Savino22}.

The LUVIT data collection and reduction was carried out separately using DOLPHOT as well and we refer the reader to the survey paper (K. M. Gilbert 2024, accepted AAS) and \citet{Gull22} for further details. We note that we only use the UV data points from the catalog.

Once the data are reduced, we apply the following quality cuts as outlined by \citet{Williams14} to each epoch; (1) signal-to-noise $<4$ , (2) \texttt{sharpness}$^2$ $<0.15,0.2$ for WFC3/UVIS, and ACS/WFC respectively, (3) \texttt{crowding} $<1.3,2.25$ for WFC3/UVIS, and ACS/WFC, and (4) any DOLPHOT FLAG $<=2$ are removed.

Figure~\ref{cmd} shows the upper part of the UV and optical color–magnitude
diagrams (CMDs) of WLM generated by the LUVIT photometry, with WLM-CB1 highlighted. We overplot the rotating ($v/v_{\mathrm{crit}} = 0.4$) MESA Isochrones and Stellar Tracks \citep[MIST;][]{choi16,dotter16} for reference. We note that the MIST isochrones have not yet been updated with in-flight characteristics of \jwst\ , which affects the bandpasses at the level of $\sim0.05$~mag. The system appears to be located on the main-sequence (MS) in each CMD, which indicates that at least one of the stars is likely a bright, blue MS star. The UV data only provides a single point, meaning without constraints on the period we do not know where in the light curve this data point falls, leading to some uncertainty in its position on the CMD.

\subsection{\jwst\ Imaging and Photometry}
The \jwst\ Resolved Stellar Populations Early Release Science (ERS) Program (DD-1334; PI D. Weisz; \citealt{Weisz23, Weisz24}) obtained F090W/F150W/F277W/F444W NIRCam imaging of a field in WLM that spatially overlaps the \hst\ imaging \citep{McQuinn24}.

The extraction of time series data points from the \jwst\ imaging required manual adjustments to the default \jwst\ pipeline. The ERS program was not formally designed as a time-series program (i.e., specifically using the time series observation mode, TSO), as it required spatial dithers to sample the point spread function, which is not permitted in TSO mode. As a result, the \jwst\ pipeline by default does not produce exposure level calibrated images needed for DOLPHOT. We therefore created them by re-running the pipeline manually with small changes. Specifically, we started with the Stage 1 \texttt{*\_rateints} data products from the Mikulski Archive for Space Telescopes (MAST). We passed into the Stage 2 imaging pipeline\footnote{\url{https://jwst-pipeline.readthedocs.io/en/latest/jwst/pipeline/calwebb\_image2.html}} (\texttt{calwebb\_image2} or \texttt{jwst.pipeline.Image2Pipeline}) to produce the fully calibrated \texttt{*\_calints} products. We stopped our reduction here, and did not produce aperture photometry, yet. Instead we first created a series of individual``pseudo-cal'' 2D images by splitting the \texttt{*\_calints} products. This choice was made so that the data is compatible with our existing pre-processing routine for DOLPHOT \citep{Dolphin00,DOLPHOT} input images. The ``pseudo-cal'' image files were supplemented by a table of integration times (FITS extension ``INT\_TIMES''), so that we retained the correct timestamps for each header. 

A reduction from this point on was performed via the DOLPHOT package for NIRCam \citep{Dolphin00,DOLPHOT, Weisz24}. 
We applied the following quality cuts to the per-epoch catalog as outlined in Section 3 of \citet{Warfield23}; (1) OBJECT\_TYPE $<=2$; (2) \texttt{sharpness}$^2$ $<0.01$ for both F090W and F150W; (3) \texttt{crowding} $<0.5$ for both F090W and F150W; and (4) any FLAG $<=2$.
While the \jwst\ photometry reaches magnitudes of $\sim 28$~ mag, we only highlight the relevant part of the upper CMD in Figure~\ref{cmd}. We summarize the time-stamps of the individual multi-band photometric observation in Table \ref{timeseries}.

\begin{deluxetable*}{ccccc}
\tablewidth{\columnwidth}
\tablecaption{ Time-stamps of individual WLM-CB1 observations.}
\tablehead{\colhead{Filter}&\colhead{MJD}&\colhead{Magnitude}&\colhead{Uncertainty}&\colhead{Systematic Uncertainty}}
\startdata
\hst\ ACS/WFC F475W & 57220.347346 &21.619& 0.005 &0.03 \\
\hst\ ACS/WFC F475W & 57220.410298 & 21.478& 0.004 &0.03 \\
\hst\ ACS/WFC F475W & 57220.479904 & 21.354& 0.004 &0.03 \\
\ldots & \ldots & \ldots & \ldots & \ldots\\
\enddata
\tablecomments{\label{timeseries} This Table is published in its entirety in the machine readable format. A portion is shown here for guidance regarding its form and content. }
\end{deluxetable*}

\section{Analysis} \label{sec:lc}
We first describe the discovery of WLM$-$CB1 in archival \hst\ data, we then determine a credible period range of WLM$-$CB1 using multi-band Lomb-Scargle (L-S) periodogram provided with \texttt{astropy} \citep{astropy:2013,astropy:2018,astropy:2022,astroML}. Next, we use \texttt{PHOEBE} \citep[PHysics Of Eclipsing BinariEs;][]{Prsa05,Prsa16,Horvat18,Conroy20,Jones20}, a mature software package that allows in-depth modeling of light and RV curves, to determine the parameters of the binary system. Below, we describe each step in more detail. 

\subsection{Discovery}
We initially apply an additional magnitude cut to the \hst\ archival data that filters for light-curves where the difference between the minimum and maximum reported magnitude in F475W and F814W is $\ge0.1$~mag. We then use multi-band L-S periodogram in astropy \citep{astropy:2013,astropy:2018,astropy:2022,astroML} to find any light-curves in the \hst\ archival data that may show periodic signal. We search over a range of periods from $0.2$ to $3$~days, with a step size of $0.001$~days. The min-max range was set by the properties of our observations, which were taken for the purposes of studying RR Lyrae. The best candidates are then visually inspected, specifically whether the periodicity was visible and similar in nature in both bands. In addition the star in this paper, we detected an three other massive or intermediate mass eclipsing short-period binary candidates in WLM, which will be reported in a future study. We choose to focus on WLM-CB1 first as it is the brightest source.

\begin{figure*}[ht!]
\includegraphics[width=18cm]{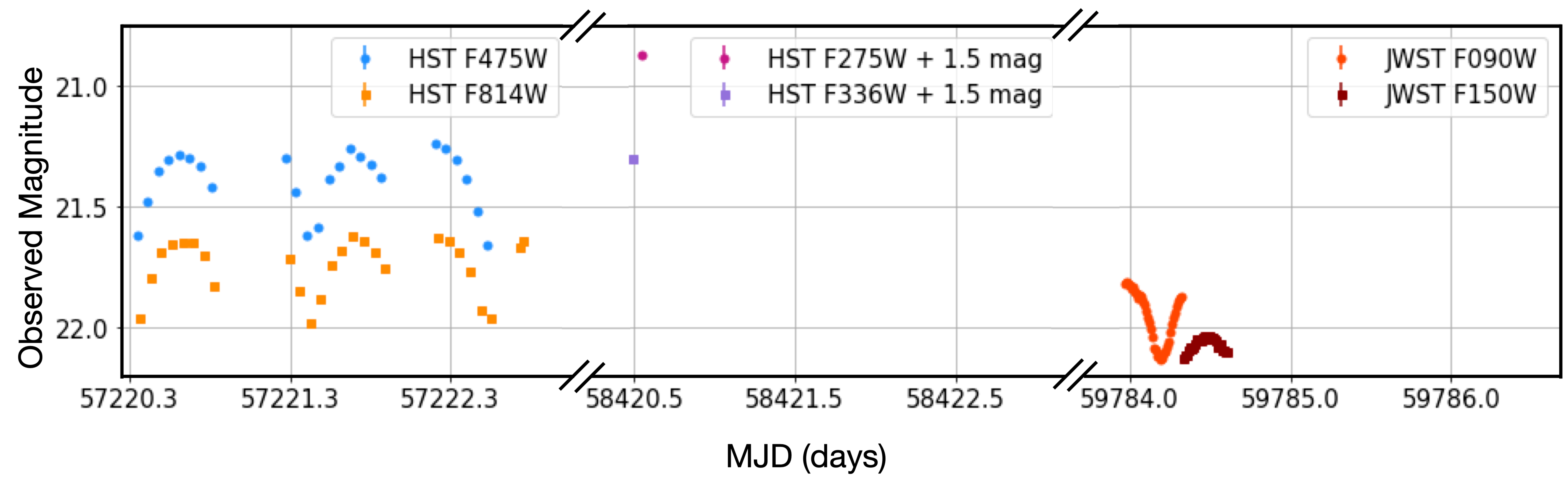} \caption{All \hst\ and \jwst\ observations of the WLM-CB1 used in this paper, shown in chronological order. Even without phase-folding, variability in the light-curves the variability in the optical and near-IR bands. A constant offset has been added to the two UV data points in order to improve the dynamic range of the y-axis for illustrative purposes.\label{rawdata}}
 \end{figure*}

Figure~\ref{rawdata} shows the multi-band light curves of WLM$-$CB1. Photometric variations are already quite clear in F475W, F814W, F090W, and F150W, indicating this is not a normal, static, upper MS star. Although our data does not completely cover the entire phase, our sampling is still clearly rich in information.
The overall shape of the light-curve suggest that the system could be a contact binary or near-contact (i.e., semi-detached) binary, hence why a quantitative analysis with \texttt{PHOEBE} is necessary to determine the exact nature of WLM-CB1.

\subsection{Period and Temperature Estimation} \label{sec:pandtest}

We use the multi-band Lomb-Scargle periodogram to estimate the period of WLM-CB1, now including the IR data points. The UV data points do not contribute to the period determination in the periodogram, due to the sparse coverage. 

For the combined F475W, F814W, F090W and F150W photometry, the L-S power spectrum yields a period of $1.0934$~days and a FWHM of $0.07$. This period is the orbital period, since the original search only using F475W and F814W yielded another peak at $0.5467$~days likely half the period. we note that the peak is broad, nearly rectangular, hence why bounds of our model grid for the period are chosen to be slightly larger than the uncertainty yielded by the FWHM. 

In the absence of spectra, we use multi-band colors to infer a range of plausible temperatures for this system. We later use this range to inform our more detailed model grid for formal fitting. We choose to refer to the hotter star as the primary and the cooler star as the secondary. Specifically, we place WLM-CB1 on static CMDs (See Figure~\ref{cmd}) using the peak magnitude for WLM-CB1 (when available). WLM-CB1 is located on the upper MS in the optical and IR CMD, close to the zero-age main sequence (ZAMS), suggesting a hot MS primary, since the primary will dominate the SED. In the UV, we shade the color-magnitude range in which the star is likely to be located, since we do not have the peak magnitude observed. While we cannot precisely place the peak magnitude, the location still suggests a hot system since it is on the MS. The secondary is unlikely to be a cool star, since the star would not appear on the MS on the near-IR CMD. We therefore assume that the stars must both be hot MS stars. Based on the color and magnitude of the peak light we estimate using the evolutionary tracks that the primary will have a temperature between $24,000$~K and $35,000$~K, which would be a late O-type or early B-type star. We allow the secondary to reach the same maximum temperature; however, we allow it to be a late B-type star, which translates to a lower bound of $~10,000$~K. Hence the temperature and mass range are quite broad, as might be expected when using only a photometric estimation. Furthermore, we can compare WLM-CB1 against mass tracks in the CMD. In each CMD, WLM-CB1 is located above the 10 \msun\ track, and close to the 25 \msun\ track in the optical and in the IR. Based on this we can infer that at least one of the stars is massive ($M_{\star}>8$\msun) , while the secondary is either an intermediate mass or massive star ($M_{\star}>3$\msun) .


\subsection{Neural Net training and Setup}

Though \texttt{PHOEBE} provides all the functionality needed to characterize our data, it is computationally expensive to embed and run it within a Markov chain Monte Carlo framework \citep[e.g.,][]{Conroy20,Ding21,Ding22}, especially when fitting a great number of models to multiple observed light curves. Therefore, following \citet{Ding21,Ding22}, we train a neural net framework to emulate \texttt{PHOEBE}, resulting in significantly reduced model generation time with minimal loss in accuracy.

Because it is not immediately clear whether WLM-CB1 is a true contact binary or a semi-detached system, we generated a model grid for each scenario in \texttt{PHOEBE}. At this point, the parameterization of \texttt{PHOEBE} does not allow for a smooth transition between detached (which includes semi-detached) and contact binary models, meaning the analysis has to be done separately for each physical scenario. We train different neural nets for each grid. 

In what follows, we describe the fitting process for the single case of the contact binary, and note that we carried out an identical process for the detached model grid. We opt for a fully connected neural network (NN), using PyTorch \citep{pytorch} and Optuna \citep{optuna_2019}. Third-light contributions (i.e., crowding from a background object, or an extra component in the system (tertiary)) are not invoked and neither are stellar spots. 



For the contact binary model grids, we have the following physical parameters: temperature of the primary ($T_{1}$), mass of the primary ($M_{1}$), temperature ratio ($T_{2}/T_{1}$), inclination ($Inc$), mass ratio ($q = M_{2}/M_{1}$), period ($P$), fillout-factor (\textit{f}) and extinction (\av). For the detached system model grid, the \texttt{PHOEBE} parameters are: temperature of the primary ($T_{1}$), mass of the primary ($M_{1}$), radius of the primary ($R_{1}$), temperature of the secondary ($T_{2}$), radius of the secondary ($R_{2}$), inclination (Inc), mass ratio (q = $M_{2}/M_{1}$), period (P), and extinction (\av). Our adopted parameter ranges are displayed in Table \ref{nn_grid}. 

\corr{The temperature range of the primary and secondary was chosen based on CMD location, as explained in Section~\ref{sec:pandtest}.} Similarly, the mass ranges are informed by the location the system in the CMD above the 10 \msun\ evolutionary track. The radius range is determined using average radii for MS stars observed in the SMC and sub-SMC galaxies. The period range was determined using the results from the periodogram, where we use the base of the peak in the L-S spectrum as our range. The inclination range is set to a lower limit of $40^\circ$, since observing a lower inclination would be unlikely given the depth of the eclipse. The mass-ratio is set at a lower cut-off of $0.2$, since observing a smaller companion than this would be unlikely, as such a low-mass would not perturb the orbit of a massive star so much. We note that the upper limit is set to $>1$ for mass-ratio and temperature ratio. Although we choose the brighter/more massive star to be the primary (i.e., subscript 1), we needed to ensure that the model can reliably recover a equal mass or equal temperature model.

\begin{deluxetable*}{cccccccccccc}[ht!]
\tablecaption{\label{nn_grid} Basic information on the \texttt{PHOEBE} model grids used for the neural network. The columns are: (1) Model type; (2) Temperature of the primary ($T_{1}$); (3) Mass of the primary ($M_{1}$); (4) Temperature Ratio ($T_{2}/T_{1}$); (5) Inclination ($Inc$);(6) Mass ratio ($q = M_{2}/M_{1}$); (7) Period ($P$); (8) fillout-factor (\textit{f}); (9) extinction (\av); (10) Temperature of the secondary ($T_{2}$); (11) Radius of the Primary ($R_{1}$); (12) Radius of the secondary ($R_{2}$) }
\tablehead{\colhead{Mode}&\colhead{$T_{1}$} &\colhead{$M_{1}$}&\colhead{$T_{2}/T_{1}$} &\colhead{$Inc$}&\colhead{$q$}&\colhead{$P$}&\colhead{\textit{f}}&\colhead{\av} &\colhead{$T_{2}$} &\colhead{$R_{1}$}&\colhead{$R_{2}$} \\
\colhead{}&\colhead{K}&\colhead{\msun}&\colhead{}&\colhead{deg}&\colhead{}&\colhead{days}&\colhead{}&\colhead{}&\colhead{K}&\colhead{\rsun}&\colhead{\rsun}\\
\colhead{(1)}&\colhead{(2)}&\colhead{(3)}&\colhead{(4)}&\colhead{(5)}&\colhead{(6)}&\colhead{(7)}&\colhead{(8)}&\colhead{(9)}&\colhead{(10)}&\colhead{(11)}&\colhead{(12)}}
\startdata
Contact & 24000-35000 & 8-20 &0.3-1.05 & 40-90 & 0.2-1.1 & 1.0-1.2 & 0.01-0.8 & 0.05-0.20 & \nodata & \nodata & \nodata\\
Detached & 20000-35000 & 8-20 & \nodata & 40-90 & 0.3-1.1 & 1.0-1.2 & & 0.05-0.20 & 15000-29000 & 3.8-7.0 & 3.8-6.5 \\
\enddata
\end{deluxetable*}

We undertake the following steps to train our neural net.
We manually add \hst\ and \jwst\ bands, which are not natively included in \texttt{PHOEBE}, using the atmosphere models of \citet{castelli04}. 
Using \texttt{PHOEBE}, we generate $\sim45,000$ light curves for each filter in phase space. The light curves are generated by randomly sampling parameters within the ranges in Table~\ref{nn_grid}. \texttt{PHOEBE} ensures that each combination of parameter yields a physical binary model. This yields a grid, where for each combination of parameters there are six associated light curves. \texttt{PHOEBE} light curves fluxes are converted to magnitudes using the \texttt{pyphot}\footnote{https://mfouesneau.github.io/pyphot/} zero-points, and parameters and light curves are normalized to improve our computational efficiency. The optimal suggested set-up yielded by tuning the neural network is a total of 5 layers; one input layer, 3 hidden layers and one output layer. The suggested activation function is ``Adam'' \citep{kingma17}, while the learning rate is $10^{-4}$ and the batchsize is set to 32.
We use 80\% of our sample to train and 20\% to validate. Our median approximation error peaks under $\sim$ $10^{-3}$ for our normalized data set. 

\begin{figure*}[ht!]
\includegraphics[width=18cm]{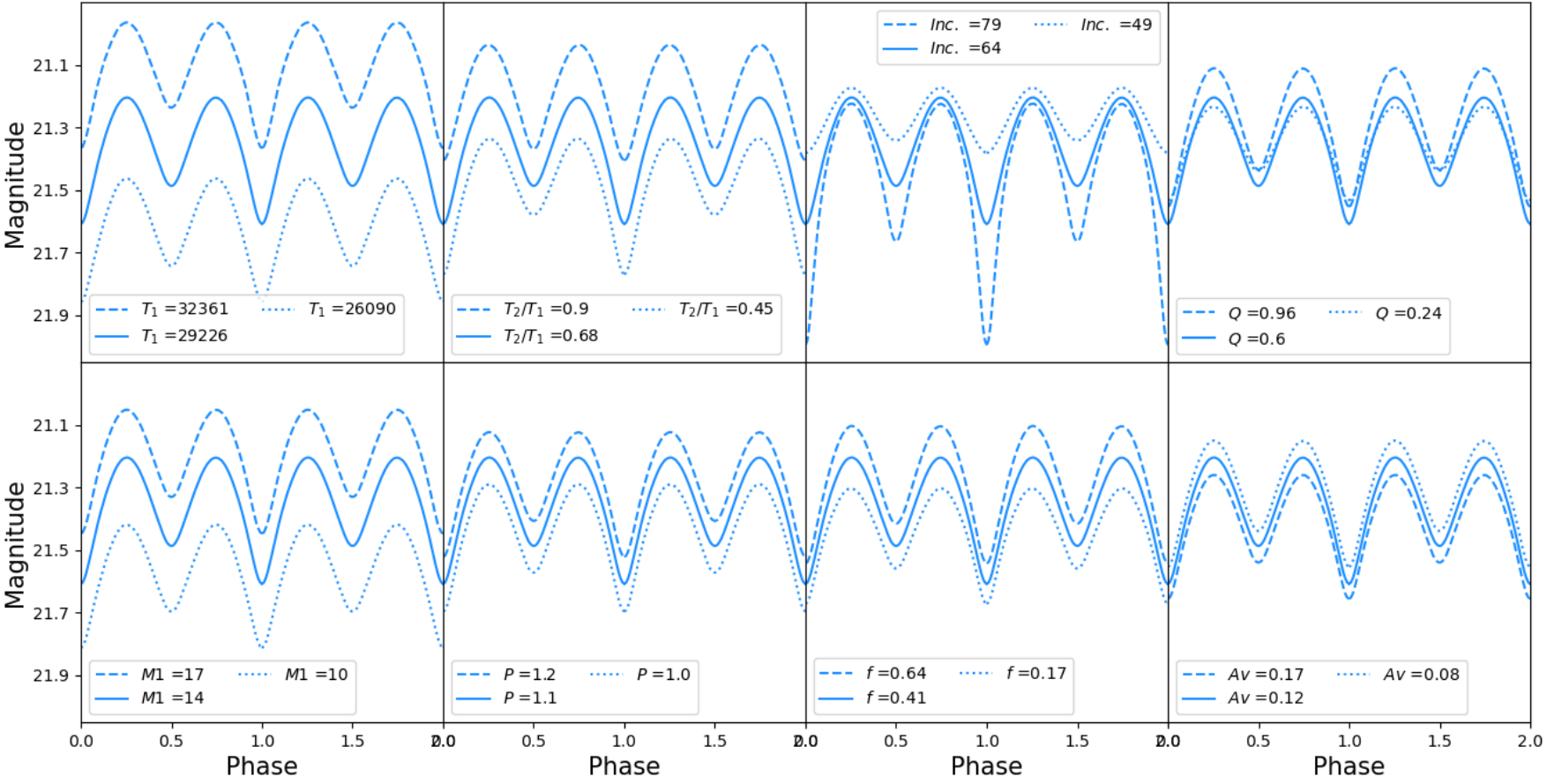} 
 \caption{Synthetic F475W light-curves generated by the neural network trained on \texttt{PHOEBE}. The solid light-curve in each panel is the baseline model. Each panel shows the effects of a singular parameter being varied on the light-curve. It is evident that the mass-ratio has the smallest effect in the light-curve and can only be constrained by high-precision photometry such \hst\ and \jwst. We note that since this is in phase space, the effects of the period on the fitting is not fully captured, as it also folds the data into phase space. 
}
 \label{poststamp}
\end{figure*}

Using the trained neural network, we present the effects of each parameter on the light curve in Figure~\ref{poststamp}. We note that the full effects pf two parameters is not fully captured by a phase-spaced single-band light-curve: Period ($P$) and Attenuation (\av\ ). The Period, additionally to the modification of the brightness of the light-curve, also will fold the data and hence have an additional temporal effect. The attenuation can only be fully captured with panchromatic light-curve since the effect on each band will be different in accordance with the dust law applied. Furthermore Figure~\ref{poststamp} shows nicely why from ground-based photometry parameters like the temperatures of each star ($T_1$ and $T_2$/$T_1$), the inclination ($Inc$) and the fill-out factor (\textit{f}) are easier to obtain. In contrast, the mass-ratio ($q$) has been notoriously difficult to obtain from light-curves due to the need for high-precision photometry, which thus far has only been possible with \hst\ or \jwst . Lastly, the mass and temperature of the primary seem to have similar effects on the single-band light curve, however, they will have different effects on the panchromatic light-curves, which allows us to disentangle the effects.

\subsection{Fitting using \texttt{UltraNest}}

Given that our light curves do not sample a full phase, and we do not have spectra, it is probable that our likelihood surface will be multi-modal in period, and perhaps other parameters. Accordingly, we opt to use a nested sampling algorithm instead of Markov Chain Monte Carlo (MCMC) methods, the former of which is generally better at sampling complex posterior distributions.

For our analysis, we use the nested sampling Monte Carlo algorithm \texttt{MLFriends} \citep{Buchner16,Buchner19} as part of the \texttt{UltraNest}~\footnote{\url{https://johannesbuchner.github.io/UltraNest/}} package \citep{Buchner21}. UltraNest uses multiple clusters and has an adjustable step sampler, which allows us to explore the entire parameter space and find the highest probability mode. To perform the fitting, we use Ultranet's reactive nested sampling implementation. Reactive nested sampling differs from a nested sampling algorithm by adding new live-points closer to the peak of the posterior mass as opposed to uniformly over the entire posterior mass. This makes the computation overall more efficient by minimizing rejection points, and does not affect the final posterior distribution. 

We use 400 live points, and, before running the algorithm, use the calibrator capabilities to obtain the optimal number of steps. Each live point takes $N_{param}\times2^{11}$ steps, so that we converge and pass the relative jump distance and acceptable fraction criteria. Furthermore, we use the ``$generate_-mixture_-random_-direction$'' step sampler. The advantage of using this ``mixed'' sampler, is that it randomly changes between region sampling and step sampling \citep{Buchner21,Buchner23}. 

\begin{figure*}[ht!]
\includegraphics[width=18cm]{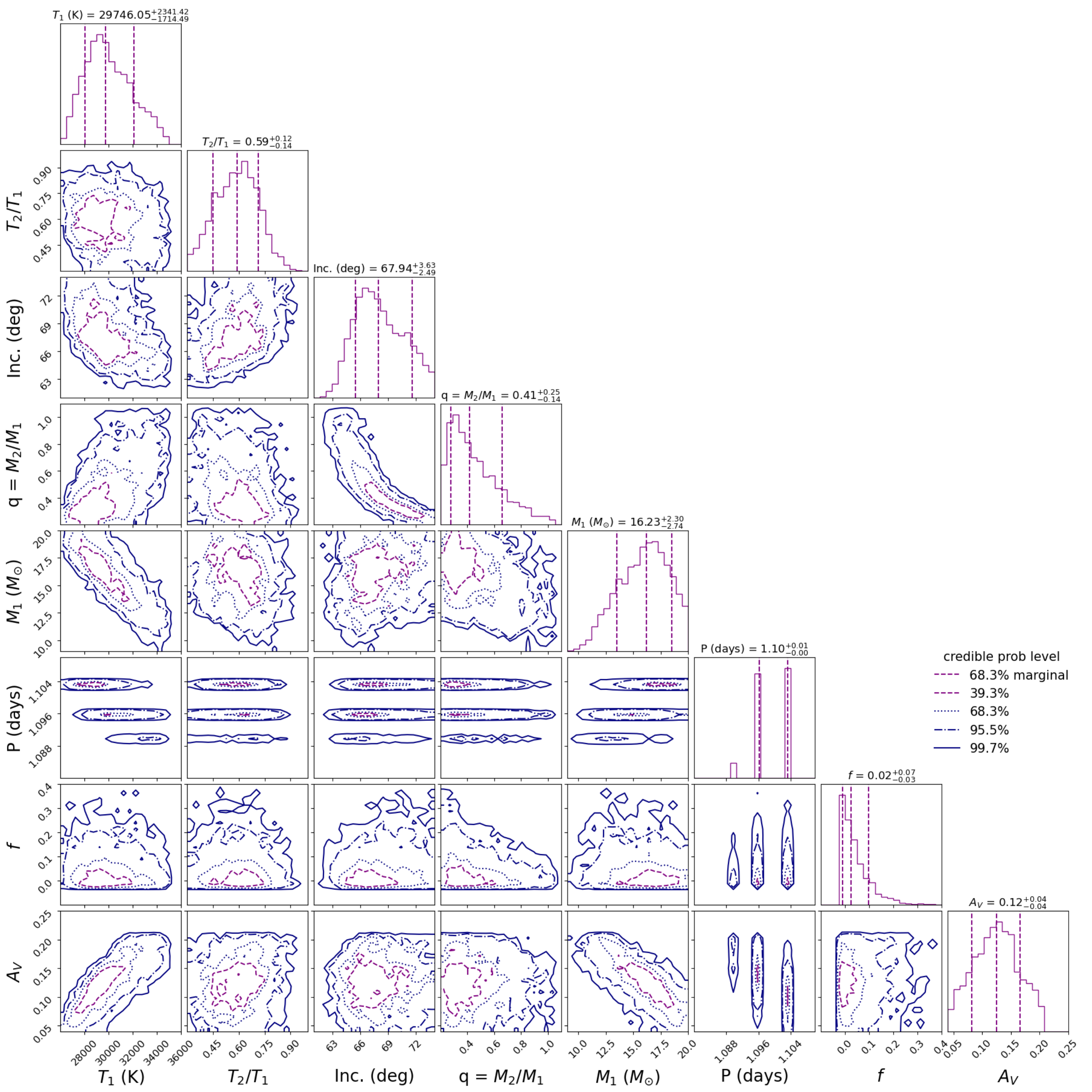} 
 \caption{A corner plot showing samples of the posterior from our ``ultranest'' fitting of WLM-CB1. For the marginalized 1D distributions, we indicate the median and 25th and 75th percentiles. The period clearly shows the mutli-modal nature of the posterior distribution. The lack of any data in the secondary dip, along with a paucity of UV data, contribute to broad constraints on temperature ratio, mass, and mass-ratio. This would be better recovered with the addition of spectroscopy. There is a correlation between temperature of the primary, mass of the primary and dust, as all parameters affect the light-curve similarly (See Figure 4).Because the fill-out factor is hitting the edge of its parameter space we consider this to be a minimal contact binary.
}
 \label{corner}
\end{figure*}

We present the resulting posterior distributions and quantiles in Figure~\ref{corner}. The strong peaks, with $0\%$ valleys for the period, reflect the multi-modality of the posterior of the period. This means that instead of a smooth probability curve, the period probability is rather concertized; very narrow peaks will set the possible values. This implies that even a slight shift in period drastically worsens the fit of model to data. 

To describe the system completely and compare it to evolutionary models, we report additional parameters. The parameters that are directly derived from our results are temperature of the secondary ($T_{2} = T_{1}\times T_{2}/T_{1}$), mass of the secondary ($M_{2} = M_{1}\times q$), and the semi-major axis ($a$). Using \texttt{PHOEBE} and the underlying atmosphere models we compute the radius of the primary ($R_{1}$), surface gravity of the primary (\logg$_1$), radius of the secondary ($R_{2}$) and surface gravity of the secondary (\logg$_2$). We present a summary of all fitted and derived parameters in Tables \ref{parameters}.

\begin{figure*}[ht!]
\includegraphics[width=17.5cm]{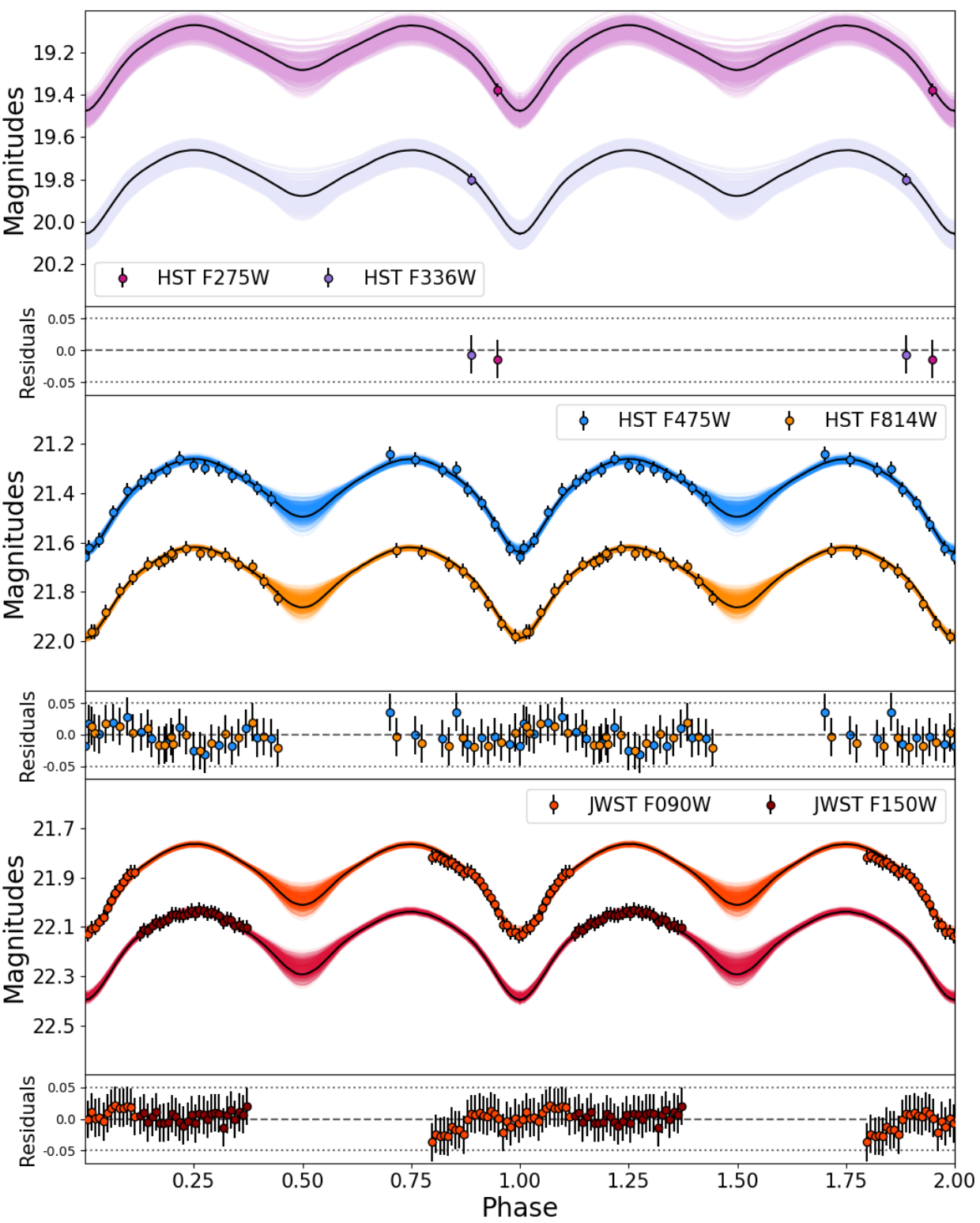}
\centering
\caption{Multi-band observed (points) and model light curves (black lines) for WLM-CB1. We also plot 300 draws from the posterior as individual colored lines. Residuals for each set of filters are plotted below the light curves. The models generally provide an excellent fit to the observed light curves. The largest uncertainties are near the minima and the UV, for which there is sparse sampling.
}\label{bestfit}
\end{figure*}

In Figure~\ref{bestfit}, we show light-curves from our data with the contact binary model light-curve generated by the median value (black) of the posterior distribution and 300 random draws from the posterior (color) overplotted . Overall, the data match the contact binary model extremely well, all data points are within the posterior bounds. The residuals are overall on the order of $<0.05$~mag, nearing our precision limit. 
The largest difference is observed in the posterior distribution around the secondary dip; however, this is mostly reflective of the uncertainties in the secondary due to lack of data in this region of phase space. We do not present the semi-detached model here, since it yields a nearly identical median light curve, which would not be noticeably different by eye.

\begin{deluxetable}{lr}
\tablewidth{\textwidth}
\tablecolumns{2} 
\tablecaption{\label{parameters} Parameters yielded by the contact binary model of WLM-CB1 binary system.}
\tablehead{ 
\colhead{Fitted parameter}\hspace{3.75cm} & \colhead{}}
\startdata
$T_{1}$ (K) & $29800_{-1700}^{+2300}$\\
$T_{2}/T_{1}$ & $0.59_{-0.14}^{+0.12}$\\
$Inc$ (degree) & $68_{-2}^{+4}$ \\
$q = M_{2}/M_{1}$ & $0.41_{-0.14}^{+0.24}$\\
$M_{1}$ ($M_{\odot}$) & $16_{-3}^{+2}$\\
$P$ (days) & $1.0961_{-0.0036}^{+0.0066}$ \\
\textit{f} & $0.02_{-0.03}^{+0.07}$\\
\av & $0.12_{-0.04}^{+0.04}$ \\
\hline
Directly derived parameter\\
\hline
$T_{2}$ (K) & $ 18000_{-5000}^{+5000}$ \\
$M_{2}$ ($M_{\odot}$) & $7_{-3}^{+5}$ \\
$a$ ($R_{\odot}$) & $12.8_{-1.1}^{+0.8}$ \\ 
\hline
\texttt{PHOEBE} derived parameter\\
\hline
$R_{1}$ ($R_{\odot}$) & 5.9 \\
\logg$_1$ (dex) & 4.11 \\
$R_{2}$ ($R_{\odot}$) & 3.9 \\
\logg$_2$ (dex) & 4.08 \\
\enddata
\end{deluxetable}

\section{Results} \label{sec:resi}
WLM-CB1 is the first observed massive sub-SMC metallicity contact binary candidate.
Our analysis suggests that WLM-CB1 is a massive contact binary consisting of a $16^{+2}_{-3}$~\msun\ star (observational primary; WLM-CB1a) and a $7^{+5}_{-3}$~\msun\ star (observational secondary; WLM-CB1b). The mass-ratio is $q = 0.41^{+0.24}_{-0.14}$. WLM-CB1a appears to be a MS star with $T_1 =29800^{+2300}_{-1700}$~(K) and \logg$_1$= 4.11, where $T_1$ is measured from the light curve and \logg$_1$\ is inferred from the constraints provided by the filling factor and mass. Similarly WLM-CB1b appears to be a MS star with $T_2= 18000^{+5000}_{-5000}$~(K) and \logg$_2$ = 4.08, which agrees well with its position on the multi-band CMDs. The system has a period of $1.0961^{+0.0066}_{-0.0036}$~days and an inferred semi-major axis of $12.8^{+0.8}_{-1.1}$~\rsun. The fill-out factor is $0.02^{+0.07}_{-0.03}$, suggesting that the system is only minimally in contact; we will further discuss this below. The extinction is \av $= 0.12 ^{+0.04}_{-0.04}$, this is only slightly higher than the overall measured foreground extinction of WLM \citep{McQuinn24}. 

In Figure~\ref{bestfit}, we show the observed light curves and those from the median values of the sampling and the posterior assuming a contact binary model. It is clear that the low sampling in the UV and the lack of coverage of the secondary eclipse drive the uncertainty. We find the data to be well-described by our contact binary model, as shown by the generally small residuals.
The fillout factor is $f=0.02^{+0.07}_{-0.03}$, which is low. As described in \citet{Bradstreet05}, a fillout factor greater than $0$ qualifies a system as a contact system. However, a system with $-0.1 \lesssim f \lesssim 0$ would alter the classification to a near contact or semi-detached system. As evident in Figure~\ref{corner}, the fillout factor is nearing the edge of the allowed range for a contact model, which would raise the concern that the model is not a contact binary. We therefore repeat the analysis assuming a (semi-) detached model. 

Quantitatively, we find little difference in the recovered physical parameters, i.e., all parameters in common are within $1-\sigma$ of one another. The sole exception is the mass-ratio, which agreed at $1.5-\sigma$. Statistically, the Bayesian evidence computed for each model was nearly identical ($logZ=-40.1 \pm 0.2$ vs $logZ=-40.8\pm 0.2$). The natural log of the Bayesian factor is less than one, indicating the current available data is not sufficiently constraining to differentiate between these physical scenarios.

The median values for the detached model suggest that WLM-CB1a fills its Roche Lobe, while WLM-CB1b is nearing Roche Lobe overflow. This would suggest that the system would be a semi-detached. However,the suggested mass-ratio yields peculiar mass for WLM-CB1b given the suggested temperature and radius. Under the assumption that WLM-CB1b is a semi-detached system, WLM-CB1b would be extremely under-luminous compared to a similar mass star. Lastly, a lower mass-ratio in the semi-detached model would push the system to be a contact binary system with the current set of derived parameters for the semi-detached model. Given these circumstances, we cannot fully rule out a semi-detached model; however, these factors indicate that a contact system seems qualitatively more likely. We further explore these scenario using more formal stellar modeling.

\section{Stellar Evolution Modeling of WLM-CB1 Using \texttt{MESA}} \label{set_mesa}

\begin{figure*}[ht!]
\includegraphics[width=17.5cm]{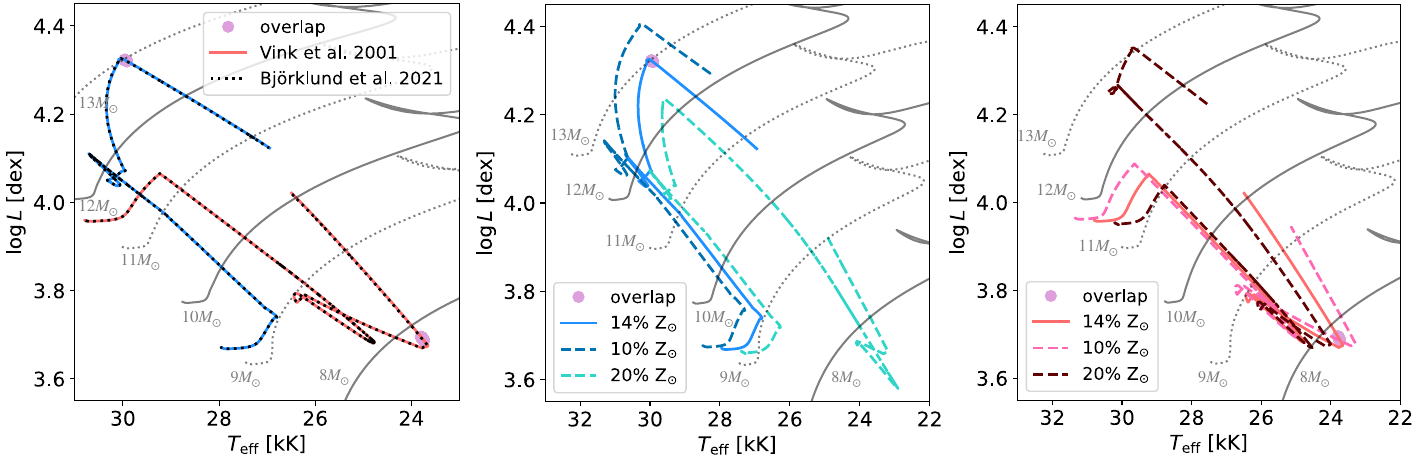}
\centering
\caption{ \corr{Evolutionary tracks yielded by MESA for our best match evolutionary path with initial primary mass ($M_{p}$) of 11.25 \msun\ for the more massive component at ZAMS, initial mass-ratio ($q_i$) of 0.8, and initial binary separation ($a_i$) of 10.75 \rsun.
\textbf{Left:} The evolutionary tracks of WLM-CB1a (blue) and WLM-CB1b (red, $M_{p}$) are shown compared to MIST single star evolutionary tracks (grey lines). The pink dots indicate the area where the current observables match. It is evident from the figure that the initial mass exchange occurs soon after the ZAMS, that the second mass exchanges also occurs on the MS and that in this particular scenario the system merges on the MS. We also overplot the evolutionary tracks computed by MESA assuming the \citet{Bjorklund21} wind prescription (black dashed line) to illustrate that in this very weak wind regime the binary interaction dominates the evolution of the stars. 
\textbf{Center and Right:} The evolutionary tracks yielded by MESA for a system with the aforementioned initial conditions ($M_{p}$=11.25 \msun , $q_i$=0.8, $a_i$=10.75 \rsun) at different metallicities of $14\%$ \zsun\ (WLM), $10\%$ \zsun (metal-poor), and $20\%$ \zsun (SMC). While the evolutionary tracks are qualitatively similar between the $14\%$ \zsun\  and $10\%$ \zsun\ track, there is a clear difference to the higher metallicity ($20\%$ \zsun) track, which experiences an additional mass transfer. The sensitivity to metallicity in binary evolution is clearly illustrated here, emphasizing the need for systematic exploration of binaries in the sub-SMC metallicity regime. It also shows that caution is necessary when extrapolating from the SMC to lower metallicity regimes.} 
}\label{hrevo}
\end{figure*}

We use the 1D stellar structure and evolution code \texttt{MESA} \citep[r12778;][]{Paxton11,Paxton13,Paxton15,Paxton18,Paxton19} to explore potential evolutionary paths of WLM-CB1. We use the assumptions in \citet{Henneco2024}, except for the metallicity ($14\%$~\zsun), which is from \citet{telford21}. We briefly summarize the \citet{Henneco2024} assumptions. Even though we use a lower metallicity, we still employ the Solar helium fraction \citep{Asplund09}. Models are assumed to be hydrostatic, and we use the \texttt{approx21} nuclear network. Stars are allowed to rotate in \texttt{MESA}'s shellular approximation and are initially synchronized with the orbital period. We impose no fixed mass-transfer efficiency. Hence, mass transfer only becomes non-conservative when the accretor star spins up to its critical rotation rate. The tidal synchronisation rate is computed through the scheme of \citet{hurley02} and the orbital angular momentum evolves through mass loss from the system and spin-orbit coupling. Orbits are assumed to be circular. We use the wind mass-loss rates derived by \corr{ \citet{Vink01} } since our models have surface temperatures above 11\,kK. 

We include mixing at several levels. Convective mixing is included via the mixing length theory \citep{bohm-vitense1958,Cox1968} with a mixing length parameter of $\alpha_{\mathrm{mlt}} = 2.0$. We employ the Ledoux criterion to determine which regions are unstable to convection. Thermohaline mixing is included with an efficiency of $\alpha_{\mathrm{th}} = 1.0$ and semi-convection with an efficiency of $\alpha_{\mathrm{sc}} = 10.0$. Rotational mixing and diffusion of angular momentum are covered by the inclusion of the Goldreich-Schubert-Fricke instability, Eddington-Sweet circulation, Spruit-Tayler dynamo, and the secular and dynamic shear instabilities. We use the Spruit-Tayler dynamo to account for additional diffusion of angular momentum. The strength of mixing of chemical elements is scaled by a factor of $f_{\mathrm{c}} = 1/30$, and the sensitivity of rotational instabilities to composition gradients, which have a stabilizing effect, is set to a value of $f_{\mu} = 0.1$. Convective boundary mixing at the interface between the convective core and radiative envelope is handled through the step-overshoot scheme. In this scheme, we extend the convective core by $0.2\,H_{\mathrm{P}}$, with $H_{\mathrm{P}}$ the pressure scale height. We compute the mass transfer rate through \texttt{MESA}'s \texttt{contact} scheme. This is a composite scheme that uses \texttt{MESA}'s \texttt{roche\_lobe} scheme for semi-detached systems and the mass-transfer scheme from \citet{Marchant16} for contact binaries. The \texttt{roche\_lobe} scheme sets the mass-transfer rate such that the donor star remains within its Roche lobe at every timestep. We stick to the original \texttt{contact} scheme by \citet{Marchant16}, that is, we do not model the tidal deformation and energy transfer in the common envelope as done in \citet{Fabry22} and \citet{Fabry23}.

\corr{We explored possible evolutionary scenarios of WLM-CB1 by first estimating a reasonable range of starting conditions for the MESA calculations (total mass, initial mass ratio and initial separation). Assuming conservative mass transfer\footnote{We find that models in this initial separation range have conservative mass transfer, but do allow for mass transfer to become non-conservative by means of the spin-up of the accretor, as mentioned above.}, we used a fixed value in the uncertainty range of the current-day total mass of WLM-CB1 as a starting point and varied the mass ratio and initial separation, with the latter limited to account only for Case-A mass transfer (see ``sweep 1'' in Table~\ref{mesa_params}). We limited the initial mass ratio to $q_{\mathrm{i}} > 0.3$ given that mass transfer is typically unstable for lower-mass-ratio systems (see, e.g., \citealt{Henneco2024} since the model assumptions are the same as in this work). We repeated this for a different value of the total mass (``sweep 2'' in Table~\ref{mesa_params}), this time restricting the initial mass ratios to $q_{\mathrm{i}} > 0.5$, with steps of 0.05 in $q_{\mathrm{i}}$ instead of 0.1 in the previous sweep. This restriction of $q_{\mathrm{i}}$ stems from the fact that no stable, long-lived (nuclear-timescale) contact binaries were found at lower $q_{\mathrm{i}}$ in the first sweep. By analyzing the results from the second sweep, we zoomed in on the most promising regions of the initial parameter space for which the models best fit our observational constraints (see ``zoom 1'', ``zoom 2'' and ``zoom 3'').} 

\corr{We emphasize that the goal of this exercise was to explore plausible evolutionary scenarios. It is, by no means, an exhaustive modeling effort to map out all possible scenarios. A more comprehensive modeling efforts would explore the initial binary parameter space in terms of the total mass, initial mass ration and initial separation in a systematic way. This approach is computationally non-trivial and beyond the scope of this work.} 

Lastly, the MESA models themselves have no constraint on whether the system is in contact or (semi-)detached; we are therefore simultaneously exploring whether the parameters match a contact or (semi-)detached point in the suggested MESA evolution.
To compare the MESA models to our observed system, we search the model grid for simultaneous match between of the observational parameters ($P$,~$q$,~$M_1$,~$M_2$,~$R_1$,~$R_2$,~$T_1$,~$T_2$,~\logg$_1$,~\logg$_2$) and a singular instance in the evolutionary models (i.e., a point in a system's evolution, where the system has the same binary properties as WLM-CB1). We use the uncertainties reported in Table \ref{parameters}, and use $0.5$ uncertainty for $R_1$, $R_2$, \logg$_1$, and \logg$_2$. \corr{We note that keeping $R_1$, $R_2$, \logg$_1$, and \logg$_2$ free does not yield any additional matches.}


\begin{deluxetable}{lcccc}
\tablewidth{\textwidth}
\tablecolumns{5} 
\tablecaption{\label{mesa_params} Parameter ranges of the different initial binary parameter sweeps and zoom-ins for the \texttt{MESA} models.}
\tablehead{ 
\colhead{name} & \colhead{no. models} & \colhead{total mass} & \colhead{$q_i$} & \colhead{$a_{i}$}\\
\colhead{} & \colhead{} & \colhead{\msun} & \colhead{} & \colhead{\rsun}}
\startdata
sweep 1 & 84        & 20.61        & 0.3 -- 0.9         & 9.16 -- 26.85      \\
sweep 2 & 225        & 23.07        & 0.5 -- 0.9         & 10.25 -- 25.60     \\
zoom 1  & 15        & 18.00 -- 21.60 & 0.8              & 11.00 -- 12.00     \\
zoom 2  & 3         & 19.80        & 0.8              & 10.25 -- 10.75     \\
zoom 3  & 3         & 20.25 -- 20.70 & 0.8              & 10.50 -- 10.75 \\    
\enddata
\end{deluxetable}

\begin{figure} [ht!]
\includegraphics[width=\columnwidth]{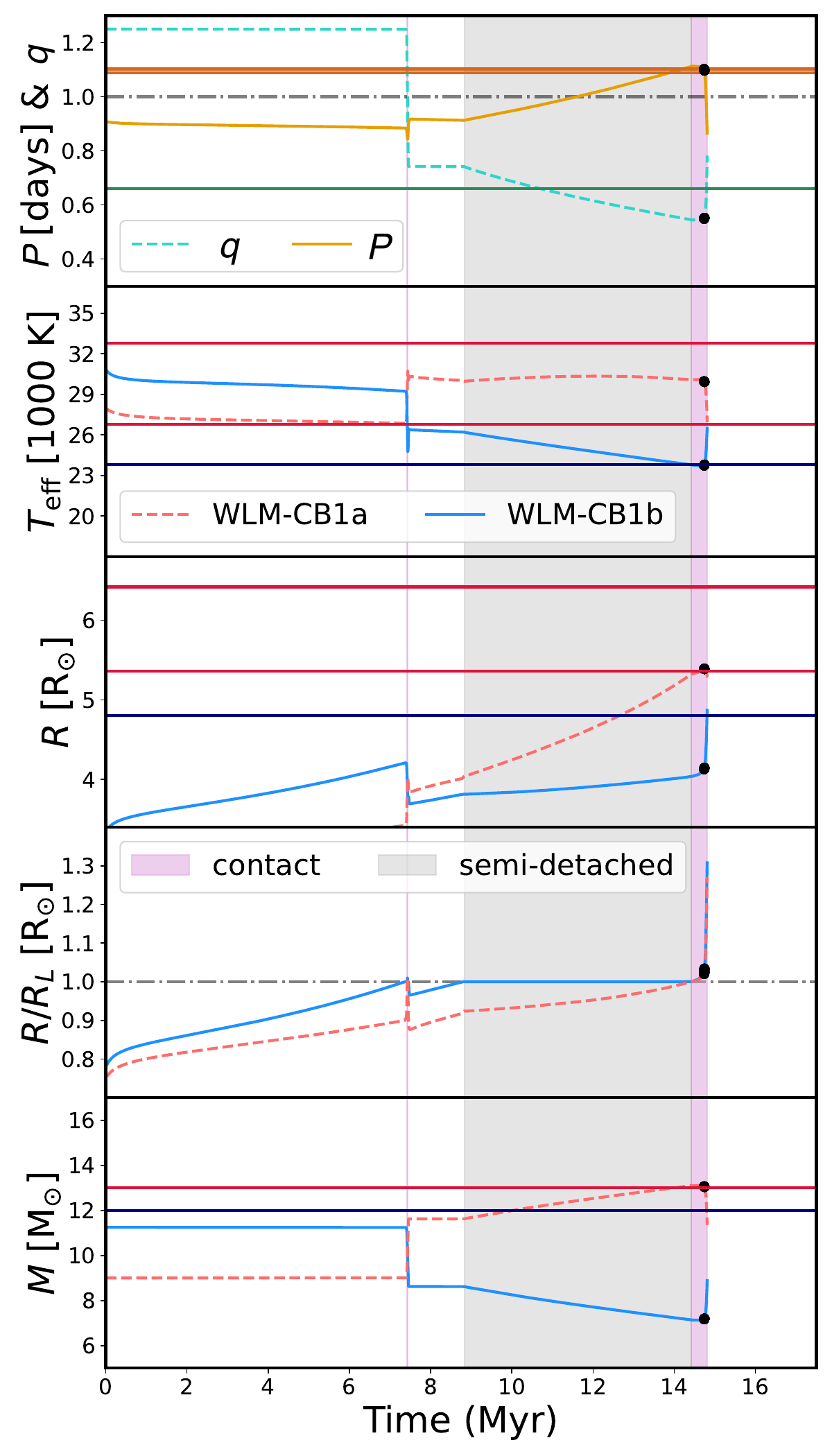}
 \caption{Best model of the evolutionary path WLM-CB1 that is consistent with current observations. The initial parameters of this model are the Mass of the primary ($Mp$) $= 11.25$~\msun, mass-ratio ($q$) $= 0.8$ and binary separation ($a_i$) = $10.75$. The instance of overlap between the models and the observational parameters are marked by black dots. The limits given by the observational uncertainties are horizontal solid lines (crimson for WLM-CB1a and navy for WLM-CB1b). As seen in the 4th row, each evolutionary path suggests that both stars have filled their Roche Lobe ($R/R_L \geq 1$ at the observed point of their evolution). The scenario suggest a previous drastic short-lived mass transfer/contact phase, where the mass-ratio is flipped (light pink vertical line). The matches between models and observation suggest that WLM-CB1 just re-entered a contact phase (pink) after a prolonged near contact phase (semi-detached, grey), hence why the definitive nature of WLM-CB1 is hard to determine. 
}
 \label{MESA}
\end{figure}

In the initial search, we are not able to find any direct matches. The closest match requires either a relaxation on the lower bound of $M_1$ from $3$\msun to $3.3$\msun or a relaxation on the upper bound of $T_2$ from $5000$~K to $5800$~K. 
The evolutionary path with the closest match suggests an initial mass ($M_{p}$) of 11 or 11.25 \msun\ for the more massive component at ZAMS, initial mass-ratio ($q_i$) of 0.8, and initial binary separation ($a_i$) of 10.75 \rsun. We show the MESA model $M_{p} = 11.25$, $q_i = 0.8$, $a_i = 10.75$ in Figure~\ref{MESA}. Interestingly, all evolutionary paths for which we find a match with WLM-CB1 paint the same picture: The two stars are born relatively close to each other and experience an initial short period of contact and extreme fully conservative mass transfer early in their evolution. This implies that WLM-CB1a was initially the less massive star and accreted mass from WLM-CB1b, until the mass ratio was reversed, so that now WLM-CB1a is the more massive star. This short-lived exchange is then followed by a prolonged near-contact phase (5-7 Myr), until ultimately the two stars once again fill their Roche-lobes and enter a contact phase once again. 

In each evolutionary path, the solution matching our observed parameters suggests that WLM-CB1 current evolutionary state is the re-entrance to the contact phase. In this scenario, this evolutionary stage could explain why, compared to other observed contact systems so far, it may have a smaller mass-ratio. Furthermore this could also explain why determining the exact nature of WLM-CB1 is difficult, as it could be in such a transitional phase. We note that, with the current period, we were not able to find a system in the definitively semi-detached phase that was able to match our observable, even with slightly relaxed constraints. 
Lastly, the two closest matches suggest that, after a contact phase, WLM-CB1 would likely merge because of orbital decay caused by mass loss through the outer Lagrange points \citep{marchant2021,Henneco2024}. 

\corr{Recently \citep{telford24}, reported that the \citet{Vink01} predicted mass-loss rates appear to be too large for sub-SMC metallicity massive stars. For a sub-set of models we tested the \citet{Bjorklund21} prescription for mass-loss rates \corr{(See comparison Fig. \ref{hrevo})}. For WLM-CB1 the binary interaction dominates the evolution, and we see no significant differences between \citet{Vink01} and \citet{Bjorklund21}, except for a slight shift in timescale ($14.81$~Myr and $14.78$~Myr). We note that in a higher mass regime, where mass-loss can be more significant, the mass-loss recipe may play a more significant role.}

\corr{We also explored the effects of changing the metallicity of our system slightly, by using the same initial conditions but varying the metallicity. The evolutionary path is qualitatively similar at lower metallicity ($\sim 10\%$, See Fig. \ref{hrevo}). At higher metallicity ($\sim 20\%$), the system has reversed its mass ratio not once but twice. This illustrates the sensitivity of binary evolution to metallicity and that systematic studies of these sub-SMC metallicity regimes are necessary to truly capture the physics in metal-poor environments \citep[e.g.,][]{Gotberg17,Klencki20}.} 

\corr{Finally, we note that while our modeling suggests a merger is likely, it is not the only plausible outcome. It is possible that this system could survive its contact phase past the main-sequence ultimately evolving into a stripped star binary system. More detailed exploration of this possibility requires improved data (e.g., more UV data points, better cadence to constrain the period, time series radial velocities, rotational velocity, mass-loss), as well as a larger model grid that includes more subtle physical effects (e.g., tidal distortion or energy transport), and a more thorough investigation of the effects sub-SMC metallicity on binary evolution would certainly be useful.}

\corr{As mentioned above,} we acknowledge that this is only a small subset of plausible physical scenarios and our point is more illustrative than conclusive. \corr{We have assumed that the system primarily experiences conservative mass transfer, due to the observational suggestions that both stars are main-sequence stars and hence experience Case A mass-transfer. In some cases (e.g., wider Case A systems, Case-B and -C systems with at least one evolved star) this mass-transfer could be non-conservative, which is a possibility not explored in this paper, but an idea we can pursue in future work.}

\begin{figure}[ht!]
\includegraphics[width=\columnwidth]{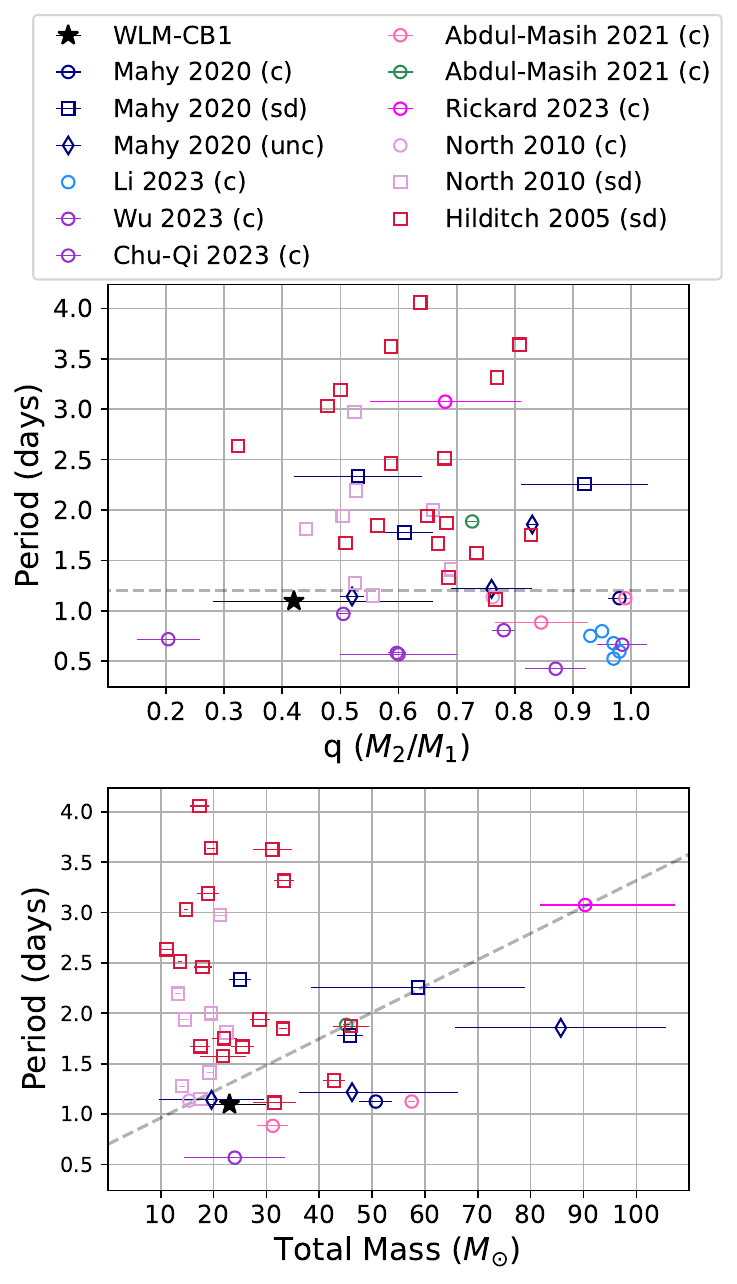}
 \caption{Overview of the observed mass ratio (q) and period (top) and of the observed total mass and period (bottom) of the contact binaries (circles) or semi-detached binaries (squares) in the LMC (blue hues) and SMC (purple/pink hues). We also plot system that are not clearly distinguished between the two (diamonds). WLM-CB1 (black star) appears to be in a regime where the transition from semi-detached to contact binary occurs based on the current observational sample. In the top panel, the gray line indicated $P=1.2$~days, to illustrate that the majority of the systems below that line are contact binaries. In the bottom panel, the gray line is a visual guide to show that if a line is draw through the marginal contact binaries in literature (V382 Cyg (MW green), $f \sim 0.1$; \citealt{AbdulMasih21}) and SSN 7 ($R_1/R_{RL}= 1.01$ and $R_1/R_{RL}= 1.03$; \citealt{Rickard23}) all observed contact system to this date fall on or below that line. We further note that the SMC appears to show a wider range of observed mass-ratios, though this may be an effect introduced by observational bias or the small sample. 
}
 \label{literature}
\end{figure}

\section{Discussion}\label{s:discussion}
\subsection{WLM-CB1: a contact binary candidate}

WLM-CB1 is the first sub-SMC metallicity contact binary candidate in the literature. Since there is currently no other confirmed sub-SMC metallicity contact binary we can only compare our system to literature examples of contact systems and semi-detached systems in the LMC and SMC \corr{and one Milky Way system} (See Figure~\ref{literature}). In this comparison, WLM-CB1 seems to be right at the boundary of semi-detached and contact systems, though the sample of known systems is small  ($\sim 50$). Qualitatively, below a $P<1.2$~days we find that out of the 20 systems in this sample, 16 are contact binaries, while 2 are semi-detached and 2 are undetermined. The frequency suggests that a contact system classification for WLM-CB1, which has a period of $P= 1.0961^{+0.0066}_{-0.0036}$~days.
If we apply the period cut and a mass-ratio cut at $q<$0.7, we find that 4 of the 6 systems that pass this cut are contact binaries. When looking at the total mass of the system vs. period, it appears that if we draw a line through V382 Cyg (\corr{Milky Way system,} $f \sim 0.1$ or $f \sim 1.1$ if $1=$ contact; \citealt{AbdulMasih21}) and SSN 7 (SMC, $R_1/R_{RL}= 1.01$ and $R_1/R_{RL}= 1.03$), which are both marginal contact systems, the majority of systems falling on or below this empirical line (including our system) are contact systems. Overall, our comparison to observations in the literature favors a contact system as opposed to a semi-detached system. However, we remain cautious as we are in a small sample regime, which is not uniform (i.e., various metallicities, masses and stellar types). 

\corr{Theoretical studies \citep[e.g.,][]{Fabry24, Menon24} have recently aimed to predict the likely ranges of parameters for contact binary systems in higher metallicity regimes (MW, LMC, SMC). While some predictions suggest that contact binaries favor equal mass configurations, it appears that initial assumptions can influence the results greatly \citep{Fabry24}. This is inline with observations of massive contact binaries \citep[e.g.,][]{AbdulMasih22} in this metallicity regime, which find unequal-mass contact binaries and do not find evidence that they are quickly evolving towards an equal-mass configuration. }

As noted we present WLM-CB1 as a contact binary candidate, since the fitting to observations does not statistical prefer a contact system vs. a near-contact/semi-detached system. We emphasize that follow-up observations as well as a more holistic analysis of the sub-SMC binary population is necessary to draw a definitive conclusion and confirm the true nature of this candidate. An increased sample of in detail studied contact binaries, which should soon be available through Vera C. Rubin Observatory's Legacy Survey of Space and Time (Rubin/LSST) \citep[e.g.,][]{Ivezic19,Street23,Hambleton23} and BlackGEM \citep[][]{Groot22}, would shed light on the distribution of binaries in the SMC and the lowest metallicity regime.

More data is required to more precisely constrain the nature of WLM-CB1. Notably the lack of temporally extended UV data and coverage of the second dip drives larger uncertainties in the fitting (e.g., Figure \ref{bestfit}. UV data points at the minima are a high priority.

\begin{figure}
\includegraphics[width=\columnwidth]{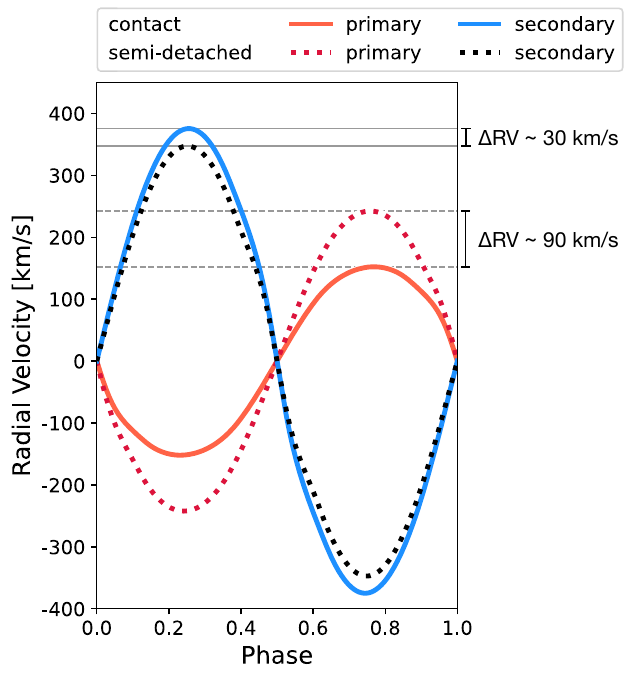}
 \caption{Synthetic Radial Velocity curves obtained by the \texttt{PHOEBE} code. We compute the expected radial velocity curves using the parameters obtained by the analysis. The solid lines represent the expected curves yielded by the contact fit, whereas the dotted lines are the expected curves yielded by the semi-detached fit. The velocity of the primary is more affected by the difference in parameters, since the difference is primarily introduces by the mass-ratio, i.e., in the semi-detached case the secondary is more massive than in the contact case, therefore the orbital velocity of the primary is larger. 
}
 \label{rvcurves}
\end{figure}

Spectroscopically, time-series radial velocity measurements of the system would constrain the masses and mass ratio to a higher certainty (See Figure~\ref{rvcurves}). The estimated $\Delta$RV of $~90$~\kms\ difference between the two scenarios should be resolvable even in a low-resolution (R$\sim$2000) spectrum. Given the inferred inclination angle from the light curves and the close distance between the two stars, there is a chance to observe the system as a spectroscopic binary. To capture the $\Delta$RV of $~30$~\kms\ difference for the secondary and disentangle the effects of both stars a medium-resolution spectra is a preferred option. Observing the spectral features of both stars, from emission lines to metal abundances, would provide further constraints on the evolutionary stage of the system. Comparison between photometry and spectroscopy will become increasingly important as we enter an era of time domain photometry and aim to study large populations of binaries yielded by these surveys. In the advent of time-series survey telescopes like the Rubin/LSST, BlackGEM, and instruments more sensitive to hot, blue objects, such as UVEX \citep{Kulkarni21} and CUBES \citep{Evans23}, WLM-CB1 showcases how we will soon be able to routinely probe binary evolution of massive stars in the lowest metallicity regimes. 

\acknowledgments
MG thanks the referee for helpful discussions, feedback or comments. MG thanks Sabrina Drammis, Jessica Lu and Jacqueline Blaum for the helpful discussion. MG acknowledges support of the UC Berkeley Cranor Fellowship and the Schweizerische Studienstiftung. 
Support for this work was provided by NASA through grants GO-15275, GO-15921, GO-16149, GO-16162, GO-16717, AR-15056, AR-16120, HST-HF2-51457.001-A, and JWST-DD-1334 from the Space Telescope Science Institute, which is operated by AURA, Inc., under NASA contract NAS5-26555.

This research used the Savio computational cluster resource provided by the Berkeley Research Computing program at the University of California, Berkeley (supported by the UC Berkeley Chancellor, Vice Chancellor for Research, and Chief Information Officer).

This work is based on photometric observations made with the NASA/ESA Hubble Space Telescope, obtained from the data archive at the Space Telescope Science Institute. STScI is operated by the Association of Universities for Research in Astronomy, Inc. under NASA contract NAS 5-26555. \corr{The HST data presented in this article can be obtained from the Mikulski Archive for Space Telescopes (MAST) at the Space Telescope Science Institute. The WLM-CB1 and WLM-CB1 UV observations analyzed can be accessed via \dataset[doi: 10.17909/v5pa-y253]{https://doi.org/10.17909/v5pa-y253} and \dataset[doi: 10.17909/xxwx-ca38]{https://doi.org/10.17909/xxwx-ca38}, respectively.}
\corr{The JWST data can also be obtained from the MAST. The observations are part of the ERS JWSTSTARS High Level Science product and can be accessed via \citep{JWSTSTARS}.}

This work made extensive use of NASA's Astrophysics Data System Bibliographic Services.

\software{\texttt{astropy} \citep{astropy:2013, astropy:2018, astropy:2022}, \texttt{astrML} \citep{astroML}, \texttt{corner} \citep{corner}, \texttt{DOLPHOT} \citep{DOLPHOT},
\texttt{matplotlib} \citep{matplotlib}, \texttt{numpy} \citep{numpy}, \texttt{MESA} \citep{Paxton11,Paxton13,Paxton15,Paxton18,Paxton19}, \texttt{Optuna} \citep{optuna_2019}, \texttt{PHOEBE} \citep{Prsa16,Horvat18,Conroy20,Jones20}, \texttt{pyphot},
\texttt{PyTorch} \citep{pytorch}, \texttt{UltraNest} \citep{Buchner21}, }

\bibliography{mybib}

\begin{thebibliography}{}
\expandafter\ifx\csname natexlab\endcsname\relax\def\natexlab#1{#1}\fi
\providecommand{\url}[1]{\href{#1}{#1}}
\providecommand{\dodoi}[1]{doi:~\href{http://doi.org/#1}{\nolinkurl{#1}}}
\providecommand{\doeprint}[1]{\href{http://ascl.net/#1}{\nolinkurl{http://ascl.net/#1}}}
\providecommand{\doarXiv}[1]{\href{https://arxiv.org/abs/#1}{\nolinkurl{https://arxiv.org/abs/#1}}}

\bibitem[{{Abbott} {et~al.}(2016){Abbott}, {Abbott}, {Abbott}, {Abernathy}, {Acernese}, {Ackley}, {Adams}, {Adams}, {Addesso}, {Adhikari}, {Adya}, {Affeldt}, {Agathos}, {Agatsuma}, {Aggarwal}, {Aguiar}, {Aiello}, {Ain}, {Ajith}, {Allen}, {Allocca}, {Altin}, {Anderson}, {Anderson}, {Arai}, {Arain}, {Araya}, {Arceneaux}, {Areeda}, {Arnaud}, {Arun}, {Ascenzi}, {Ashton}, {Ast}, {Aston}, {Astone}, {Aufmuth}, {Aulbert}, {Babak}, {Bacon}, {Bader}, {Baker}, {Baldaccini}, {Ballardin}, {Ballmer}, {Barayoga}, {Barclay}, {Barish}, {Barker}, {Barone}, {Barr}, {Barsotti}, {Barsuglia}, {Barta}, {Bartlett}, {Barton}, {Bartos}, {Bassiri}, {Basti}, {Batch}, {Baune}, {Bavigadda}, {Bazzan}, {Behnke}, {Bejger}, {Belczynski}, {Bell}, {Bell}, {Berger}, {Bergman}, {Bergmann}, {Berry}, {Bersanetti}, {Bertolini}, {Betzwieser}, {Bhagwat}, {Bhandare}, {Bilenko}, {Billingsley}, {Birch}, {Birney}, {Birnholtz}, {Biscans}, {Bisht}, {Bitossi}, {Biwer}, {Bizouard}, {Blackburn}, {Blair}, {Blair}, {Blair}, {Bloemen}, {Bock}, {Bodiya}, {Boer},
  {Bogaert}, {Bogan}, {Bohe}, {Bojtos}, {Bond}, {Bondu}, {Bonnand}, {Boom}, {Bork}, {Boschi}, {Bose}, {Bouffanais}, {Bozzi}, {Bradaschia}, {Brady}, {Braginsky}, {Branchesi}, {Brau}, {Briant}, {Brillet}, {Brinkmann}, {Brisson}, {Brockill}, {Brooks}, {Brown}, {Brown}, {Brown}, {Buchanan}, {Buikema}, {Bulik}, {Bulten}, {Buonanno}, {Buskulic}, {Buy}, {Byer}, {Cabero}, {Cadonati}, {Cagnoli}, {Cahillane}, {Bustillo}, {Callister}, {Calloni}, {Camp}, {Cannon}, {Cao}, {Capano}, {Capocasa}, {Carbognani}, {Caride}, {Casanueva Diaz}, {Casentini}, {Caudill}, {Cavagli{\`a}}, {Cavalier}, {Cavalieri}, {Cella}, {Cepeda}, {Baiardi}, {Cerretani}, {Cesarini}, {Chakraborty}, {Chalermsongsak}, {Chamberlin}, {Chan}, {Chao}, {Charlton}, {Chassande-Mottin}, {Chen}, {Chen}, {Cheng}, {Chincarini}, {Chiummo}, {Cho}, {Cho}, {Chow}, {Christensen}, {Chu}, {Chua}, {Chung}, {Ciani}, {Clara}, {Clark}, {Cleva}, {Coccia}, {Cohadon}, {Colla}, {Collette}, {Cominsky}, {Constancio}, {Conte}, {Conti}, {Cook}, {Corbitt}, {Cornish}, {Corsi},
  {Cortese}, {Costa}, {Coughlin}, {Coughlin}, {Coulon}, {Countryman}, {Couvares}, {Cowan}, {Coward}, {Cowart}, {Coyne}, {Coyne}, {Craig}, {Creighton}, {Creighton}, {Cripe}, {Crowder}, {Cruise}, {Cumming}, {Cunningham}, {Cuoco}, {Dal Canton}, {Danilishin}, {D'Antonio}, {Danzmann}, {Darman}, {Da Silva Costa}, {Dattilo}, {Dave}, {Daveloza}, {Davier}, {Davies}, {Daw}, {Day}, {De}, {DeBra}, {Debreczeni}, {Degallaix}, {De Laurentis}, {Del{\'e}glise}, {Del Pozzo}, {Denker}, {Dent}, {Dereli}, {Dergachev}, {DeRosa}, {De Rosa}, {DeSalvo}, {Dhurandhar}, {D{\'\i}az}, {Di Fiore}, {Di Giovanni}, {Di Lieto}, {Di Pace}, {Di Palma}, {Di Virgilio}, {Dojcinoski}, {Dolique}, {Donovan}, {Dooley}, {Doravari}, {Douglas}, {Downes}, {Drago}, {Drever}, {Driggers}, {Du}, {Ducrot}, {Dwyer}, {Edo}, {Edwards}, {Effler}, {Eggenstein}, {Ehrens}, {Eichholz}, {Eikenberry}, {Engels}, {Essick}, {Etzel}, {Evans}, {Evans}, {Everett}, {Factourovich}, {Fafone}, {Fair}, {Fairhurst}, {Fan}, {Fang}, {Farinon}, {Farr}, {Farr}, {Favata}, {Fays},
  {Fehrmann}, {Fejer}, {Feldbaum}, {Ferrante}, {Ferreira}, {Ferrini}, {Fidecaro}, {Finn}, {Fiori}, {Fiorucci}, {Fisher}, {Flaminio}, {Fletcher}, {Fong}, {Fournier}, {Franco}, {Frasca}, {Frasconi}, {Frede}, {Frei}, {Freise}, {Frey}, {Frey}, {Fricke}, {Fritschel}, {Frolov}, {Fulda}, {Fyffe}, {Gabbard}, {Gair}, {Gammaitoni}, {Gaonkar}, {Garufi}, {Gatto}, {Gaur}, {Gehrels}, {Gemme}, {Gendre}, {Genin}, {Gennai}, {George}, {Gergely}, {Germain}, {Ghosh}, {Ghosh}, {Ghosh}, {Giaime}, {Giardina}, {Giazotto}, {Gill}, {Glaefke}, {Gleason}, {Goetz}, {Goetz}, {Gondan}, {Gonz{\'a}lez}, {Castro}, {Gopakumar}, {Gordon}, {Gorodetsky}, {Gossan}, {Gosselin}, {Gouaty}, {Graef}, {Graff}, {Granata}, {Grant}, {Gras}, {Gray}, {Greco}, {Green}, {Greenhalgh}, {Groot}, {Grote}, {Grunewald}, {Guidi}, {Guo}, {Gupta}, {Gupta}, {Gushwa}, {Gustafson}, {Gustafson}, {Hacker}, {Hall}, {Hall}, {Hammond}, {Haney}, {Hanke}, {Hanks}, {Hanna}, {Hannam}, {Hanson}, {Hardwick}, {Harms}, {Harry}, {Harry}, {Hart}, {Hartman}, {Haster}, {Haughian},
  {Healy}, {Heefner}, {Heidmann}, {Heintze}, {Heinzel}, {Heitmann}, {Hello}, {Hemming}, {Hendry}, {Heng}, {Hennig}, {Heptonstall}, {Heurs}, {Hild}, {Hoak}, {Hodge}, {Hofman}, {Hollitt}, {Holt}, {Holz}, {Hopkins}, {Hosken}, {Hough}, {Houston}, {Howell}, {Hu}, {Huang}, {Huerta}, {Huet}, {Hughey}, {Husa}, {Huttner}, {Huynh-Dinh}, {Idrisy}, {Indik}, {Ingram}, {Inta}, {Isa}, {Isac}, {Isi}, {Islas}, {Isogai}, {Iyer}, {Izumi}, {Jacobson}, {Jacqmin}, {Jang}, {Jani}, {Jaranowski}, {Jawahar}, {Jim{\'e}nez-Forteza}, {Johnson}, {Johnson-McDaniel}, {Jones}, {Jones}, {Jonker}, {Ju}, {Haris}, {Kalaghatgi}, {Kalogera}, {Kandhasamy}, {Kang}, {Kanner}, {Karki}, {Kasprzack}, {Katsavounidis}, {Katzman}, {Kaufer}, {Kaur}, {Kawabe}, {Kawazoe}, {K{\'e}f{\'e}lian}, {Kehl}, {Keitel}, {Kelley}, {Kells}, {Kennedy}, {Keppel}, {Key}, {Khalaidovski}, {Khalili}, {Khan}, {Khan}, {Khan}, {Khazanov}, {Kijbunchoo}, {Kim}, {Kim}, {Kim}, {Kim}, {Kim}, {Kim}, {King}, {King}, {Kinzel}, {Kissel}, {Kleybolte}, {Klimenko}, {Koehlenbeck}, {Kokeyama},
  {Koley}, {Kondrashov}, {Kontos}, {Koranda}, {Korobko}, {Korth}, {Kowalska}, {Kozak}, {Kringel}, {Krishnan}, {Kr{\'o}lak}, {Krueger}, {Kuehn}, {Kumar}, {Kumar}, {Kuo}, {Kutynia}, {Kwee}, {Lackey}, {Landry}, {Lange}, {Lantz}, {Lasky}, {Lazzarini}, {Lazzaro}, {Leaci}, {Leavey}, {Lebigot}, {Lee}, {Lee}, {Lee}, {Lee}, {Lenon}, {Leonardi}, {Leong}, {Leroy}, {Letendre}, {Levin}, {Levine}, {Li}, {Libson}, {Littenberg}, {Lockerbie}, {Logue}, {Lombardi}, {London}, {Lord}, {Lorenzini}, {Loriette}, {Lormand}, {Losurdo}, {Lough}, {Lousto}, {Lovelace}, {L{\"u}ck}, {Lundgren}, {Luo}, {Lynch}, {Ma}, {MacDonald}, {Machenschalk}, {MacInnis}, {Macleod}, {Maga{\~n}a-Sandoval}, {Magee}, {Mageswaran}, {Majorana}, {Maksimovic}, {Malvezzi}, {Man}, {Mandel}, {Mandic}, {Mangano}, {Mansell}, {Manske}, {Mantovani}, {Marchesoni}, {Marion}, {M{\'a}rka}, {M{\'a}rka}, {Markosyan}, {Maros}, {Martelli}, {Martellini}, {Martin}, {Martin}, {Martynov}, {Marx}, {Mason}, {Masserot}, {Massinger}, {Masso-Reid}, {Matichard}, {Matone}, {Mavalvala},
  {Mazumder}, {Mazzolo}, {McCarthy}, {McClelland}, {McCormick}, {McGuire}, {McIntyre}, {McIver}, {McManus}, {McWilliams}, {Meacher}, {Meadors}, {Meidam}, {Melatos}, {Mendell}, {Mendoza-Gandara}, {Mercer}, {Merilh}, {Merzougui}, {Meshkov}, {Messenger}, {Messick}, {Meyers}, {Mezzani}, {Miao}, {Michel}, {Middleton}, {Mikhailov}, {Milano}, {Miller}, {Millhouse}, {Minenkov}, {Ming}, {Mirshekari}, {Mishra}, {Mitra}, {Mitrofanov}, {Mitselmakher}, {Mittleman}, {Moggi}, {Mohan}, {Mohapatra}, {Montani}, {Moore}, {Moore}, {Moraru}, {Moreno}, {Morriss}, {Mossavi}, {Mours}, {Mow-Lowry}, {Mueller}, {Mueller}, {Muir}, {Mukherjee}, {Mukherjee}, {Mukherjee}, {Mukund}, {Mullavey}, {Munch}, {Murphy}, {Murray}, {Mytidis}, {Nardecchia}, {Naticchioni}, {Nayak}, {Necula}, {Nedkova}, {Nelemans}, {Neri}, {Neunzert}, {Newton}, {Nguyen}, {Nielsen}, {Nissanke}, {Nitz}, {Nocera}, {Nolting}, {Normandin}, {Nuttall}, {Oberling}, {Ochsner}, {O'Dell}, {Oelker}, {Ogin}, {Oh}, {Oh}, {Ohme}, {Oliver}, {Oppermann}, {Oram}, {O'Reilly},
  {O'Shaughnessy}, {Ott}, {Ottaway}, {Ottens}, {Overmier}, {Owen}, {Pai}, {Pai}, {Palamos}, {Palashov}, {Palomba}, {Pal-Singh}, {Pan}, {Pan}, {Pankow}, {Pannarale}, {Pant}, {Paoletti}, {Paoli}, {Papa}, {Paris}, {Parker}, {Pascucci}, {Pasqualetti}, {Passaquieti}, {Passuello}, {Patricelli}, {Patrick}, {Pearlstone}, {Pedraza}, {Pedurand}, {Pekowsky}, {Pele}, {Penn}, {Perreca}, {Pfeiffer}, {Phelps}, {Piccinni}, {Pichot}, {Pickenpack}, {Piergiovanni}, {Pierro}, {Pillant}, {Pinard}, {Pinto}, {Pitkin}, {Poeld}, {Poggiani}, {Popolizio}, {Post}, {Powell}, {Prasad}, {Predoi}, {Premachandra}, {Prestegard}, {Price}, {Prijatelj}, {Principe}, {Privitera}, {Prix}, {Prodi}, {Prokhorov}, {Puncken}, {Punturo}, {Puppo}, {P{\"u}rrer}, {Qi}, {Qin}, {Quetschke}, {Quintero}, {Quitzow-James}, {Raab}, {Rabeling}, {Radkins}, {Raffai}, {Raja}, {Rakhmanov}, {Ramet}, {Rapagnani}, {Raymond}, {Razzano}, {Re}, {Read}, {Reed}, {Regimbau}, {Rei}, {Reid}, {Reitze}, {Rew}, {Reyes}, {Ricci}, {Riles}, {Robertson}, {Robie}, {Robinet}, {Rocchi},
  {Rolland}, {Rollins}, {Roma}, {Romano}, {Romano}, {Romanov}, {Romie}, {Rosi{\'n}ska}, {Rowan}, {R{\"u}diger}, {Ruggi}, {Ryan}, {Sachdev}, {Sadecki}, {Sadeghian}, {Salconi}, {Saleem}, {Salemi}, {Samajdar}, {Sammut}, {Sampson}, {Sanchez}, {Sandberg}, {Sandeen}, {Sanders}, {Sanders}, {Sassolas}, {Sathyaprakash}, {Saulson}, {Sauter}, {Savage}, {Sawadsky}, {Schale}, {Schilling}, {Schmidt}, {Schmidt}, {Schnabel}, {Schofield}, {Sch{\"o}nbeck}, {Schreiber}, {Schuette}, {Schutz}, {Scott}, {Scott}, {Sellers}, {Sengupta}, {Sentenac}, {Sequino}, {Sergeev}, {Serna}, {Setyawati}, {Sevigny}, {Shaddock}, {Shaffer}, {Shah}, {Shahriar}, {Shaltev}, {Shao}, {Shapiro}, {Shawhan}, {Sheperd}, {Shoemaker}, {Shoemaker}, {Siellez}, {Siemens}, {Sigg}, {Silva}, {Simakov}, {Singer}, {Singer}, {Singh}, {Singh}, {Singhal}, {Sintes}, {Slagmolen}, {Smith}, {Smith}, {Smith}, {Smith}, {Son}, {Sorazu}, {Sorrentino}, {Souradeep}, {Srivastava}, {Staley}, {Steinke}, {Steinlechner}, {Steinlechner}, {Steinmeyer}, {Stephens}, {Stevenson}, {Stone},
  {Strain}, {Straniero}, {Stratta}, {Strauss}, {Strigin}, {Sturani}, {Stuver}, {Summerscales}, {Sun}, {Sutton}, {Swinkels}, {Szczepa{\'n}czyk}, {Tacca}, {Talukder}, {Tanner}, {T{\'a}pai}, {Tarabrin}, {Taracchini}, {Taylor}, {Theeg}, {Thirugnanasambandam}, {Thomas}, {Thomas}, {Thomas}, {Thorne}, {Thorne}, {Thrane}, {Tiwari}, {Tiwari}, {Tokmakov}, {Tomlinson}, {Tonelli}, {Torres}, {Torrie}, {T{\"o}yr{\"a}}, {Travasso}, {Traylor}, {Trifir{\`o}}, {Tringali}, {Trozzo}, {Tse}, {Turconi}, {Tuyenbayev}, {Ugolini}, {Unnikrishnan}, {Urban}, {Usman}, {Vahlbruch}, {Vajente}, {Valdes}, {Vallisneri}, {van Bakel}, {van Beuzekom}, {van den Brand}, {Van Den Broeck}, {Vander-Hyde}, {van der Schaaf}, {van Heijningen}, {van Veggel}, {Vardaro}, {Vass}, {Vas{\'u}th}, {Vaulin}, {Vecchio}, {Vedovato}, {Veitch}, {Veitch}, {Venkateswara}, {Verkindt}, {Vetrano}, {Vicer{\'e}}, {Vinciguerra}, {Vine}, {Vinet}, {Vitale}, {Vo}, {Vocca}, {Vorvick}, {Voss}, {Vousden}, {Vyatchanin}, {Wade}, {Wade}, {Wade}, {Waldman}, {Walker}, {Wallace},
  {Walsh}, {Wang}, {Wang}, {Wang}, {Wang}, {Wang}, {Ward}, {Ward}, {Warner}, {Was}, {Weaver}, {Wei}, {Weinert}, {Weinstein}, {Weiss}, {Welborn}, {Wen}, {We{\ss}els}, {Westphal}, {Wette}, {Whelan}, {Whitcomb}, {White}, {Whiting}, {Wiesner}, {Wilkinson}, {Willems}, {Williams}, {Williams}, {Williamson}, {Willis}, {Willke}, {Wimmer}, {Winkelmann}, {Winkler}, {Wipf}, {Wiseman}, {Wittel}, {Woan}, {Worden}, {Wright}, {Wu}, {Yablon}, {Yakushin}, {Yam}, {Yamamoto}, {Yancey}, {Yap}, {Yu}, {Yvert}, {Zadro{\.Z}ny}, {Zangrando}, {Zanolin}, {Zendri}, {Zevin}, {Zhang}, {Zhang}, {Zhang}, {Zhang}, {Zhao}, {Zhou}, {Zhou}, {Zhu}, {Zucker}, {Zuraw}, {Zweizig}, {LIGO Scientific Collaboration}, \& {Virgo Collaboration}}]{Abbott16}
{Abbott}, B.~P., {Abbott}, R., {Abbott}, T.~D., {et~al.} 2016, \prl, 116, 061102, \dodoi{10.1103/PhysRevLett.116.061102}

\bibitem[{{Abdul-Masih} {et~al.}(2022){Abdul-Masih}, {Escorza}, {Menon}, {Mahy}, \& {Marchant}}]{AbdulMasih22}
{Abdul-Masih}, M., {Escorza}, A., {Menon}, A., {Mahy}, L., \& {Marchant}, P. 2022, \aap, 666, A18, \dodoi{10.1051/0004-6361/202244148}

\bibitem[{{Abdul-Masih} {et~al.}(2019){Abdul-Masih}, {Sana}, {Sundqvist}, {Mahy}, {Menon}, {Almeida}, {De Koter}, {de Mink}, {Justham}, {Langer}, {Puls}, {Shenar}, \& {Tramper}}]{Abdul-Masih19}
{Abdul-Masih}, M., {Sana}, H., {Sundqvist}, J., {et~al.} 2019, \apj, 880, 115, \dodoi{10.3847/1538-4357/ab24d4}

\bibitem[{{Abdul-Masih} {et~al.}(2021){Abdul-Masih}, {Sana}, {Hawcroft}, {Almeida}, {Brands}, {de Mink}, {Justham}, {Langer}, {Mahy}, {Marchant}, {Menon}, {Puls}, \& {Sundqvist}}]{AbdulMasih21}
{Abdul-Masih}, M., {Sana}, H., {Hawcroft}, C., {et~al.} 2021, \aap, 651, A96, \dodoi{10.1051/0004-6361/202040195}

\bibitem[{Akiba {et~al.}(2019)Akiba, Sano, Yanase, Ohta, \& Koyama}]{optuna_2019}
Akiba, T., Sano, S., Yanase, T., Ohta, T., \& Koyama, M. 2019, in Proceedings of the 25th {ACM} {SIGKDD} International Conference on Knowledge Discovery and Data Mining

\bibitem[{{Albers} {et~al.}(2019){Albers}, {Weisz}, {Cole}, {Dolphin}, {Skillman}, {Williams}, {Boylan-Kolchin}, {Bullock}, {Dalcanton}, {Hopkins}, {Leaman}, {McConnachie}, {Vogelsberger}, \& {Wetzel}}]{Albers19}
{Albers}, S.~M., {Weisz}, D.~R., {Cole}, A.~A., {et~al.} 2019, \mnras, 490, 5538, \dodoi{10.1093/mnras/stz2903}

\bibitem[{{Almeida} {et~al.}(2015){Almeida}, {Sana}, {de Mink}, {Tramper}, {Soszy{\'n}ski}, {Langer}, {Barb{\'a}}, {Cantiello}, {Damineli}, {de Koter}, {Garcia}, {Gr{\"a}fener}, {Herrero}, {Howarth}, {Ma{\'\i}z Apell{\'a}niz}, {Norman}, {Ram{\'\i}rez-Agudelo}, \& {Vink}}]{almeida15}
{Almeida}, L.~A., {Sana}, H., {de Mink}, S.~E., {et~al.} 2015, \apj, 812, 102, \dodoi{10.1088/0004-637X/812/2/102}

\bibitem[{{Asplund} {et~al.}(2009){Asplund}, {Grevesse}, {Sauval}, \& {Scott}}]{Asplund09}
{Asplund}, M., {Grevesse}, N., {Sauval}, A.~J., \& {Scott}, P. 2009, \araa, 47, 481, \dodoi{10.1146/annurev.astro.46.060407.145222}

\bibitem[{{Astropy Collaboration} {et~al.}(2013){Astropy Collaboration}, {Robitaille}, {Tollerud}, {Greenfield}, {Droettboom}, {Bray}, {Aldcroft}, {Davis}, {Ginsburg}, {Price-Whelan}, {Kerzendorf}, {Conley}, {Crighton}, {Barbary}, {Muna}, {Ferguson}, {Grollier}, {Parikh}, {Nair}, {Unther}, {Deil}, {Woillez}, {Conseil}, {Kramer}, {Turner}, {Singer}, {Fox}, {Weaver}, {Zabalza}, {Edwards}, {Azalee Bostroem}, {Burke}, {Casey}, {Crawford}, {Dencheva}, {Ely}, {Jenness}, {Labrie}, {Lim}, {Pierfederici}, {Pontzen}, {Ptak}, {Refsdal}, {Servillat}, \& {Streicher}}]{astropy:2013}
{Astropy Collaboration}, {Robitaille}, T.~P., {Tollerud}, E.~J., {et~al.} 2013, \aap, 558, A33, \dodoi{10.1051/0004-6361/201322068}

\bibitem[{{Astropy Collaboration} {et~al.}(2018){Astropy Collaboration}, {Price-Whelan}, {Sip{\H{o}}cz}, {G{\"u}nther}, {Lim}, {Crawford}, {Conseil}, {Shupe}, {Craig}, {Dencheva}, {Ginsburg}, {Vand erPlas}, {Bradley}, {P{\'e}rez-Su{\'a}rez}, {de Val-Borro}, {Aldcroft}, {Cruz}, {Robitaille}, {Tollerud}, {Ardelean}, {Babej}, {Bach}, {Bachetti}, {Bakanov}, {Bamford}, {Barentsen}, {Barmby}, {Baumbach}, {Berry}, {Biscani}, {Boquien}, {Bostroem}, {Bouma}, {Brammer}, {Bray}, {Breytenbach}, {Buddelmeijer}, {Burke}, {Calderone}, {Cano Rodr{\'\i}guez}, {Cara}, {Cardoso}, {Cheedella}, {Copin}, {Corrales}, {Crichton}, {D'Avella}, {Deil}, {Depagne}, {Dietrich}, {Donath}, {Droettboom}, {Earl}, {Erben}, {Fabbro}, {Ferreira}, {Finethy}, {Fox}, {Garrison}, {Gibbons}, {Goldstein}, {Gommers}, {Greco}, {Greenfield}, {Groener}, {Grollier}, {Hagen}, {Hirst}, {Homeier}, {Horton}, {Hosseinzadeh}, {Hu}, {Hunkeler}, {Ivezi{\'c}}, {Jain}, {Jenness}, {Kanarek}, {Kendrew}, {Kern}, {Kerzendorf}, {Khvalko}, {King}, {Kirkby}, {Kulkarni},
  {Kumar}, {Lee}, {Lenz}, {Littlefair}, {Ma}, {Macleod}, {Mastropietro}, {McCully}, {Montagnac}, {Morris}, {Mueller}, {Mumford}, {Muna}, {Murphy}, {Nelson}, {Nguyen}, {Ninan}, {N{\"o}the}, {Ogaz}, {Oh}, {Parejko}, {Parley}, {Pascual}, {Patil}, {Patil}, {Plunkett}, {Prochaska}, {Rastogi}, {Reddy Janga}, {Sabater}, {Sakurikar}, {Seifert}, {Sherbert}, {Sherwood-Taylor}, {Shih}, {Sick}, {Silbiger}, {Singanamalla}, {Singer}, {Sladen}, {Sooley}, {Sornarajah}, {Streicher}, {Teuben}, {Thomas}, {Tremblay}, {Turner}, {Terr{\'o}n}, {van Kerkwijk}, {de la Vega}, {Watkins}, {Weaver}, {Whitmore}, {Woillez}, {Zabalza}, \& {Astropy Contributors}}]{astropy:2018}
{Astropy Collaboration}, {Price-Whelan}, A.~M., {Sip{\H{o}}cz}, B.~M., {et~al.} 2018, \aj, 156, 123, \dodoi{10.3847/1538-3881/aabc4f}

\bibitem[{{Astropy Collaboration} {et~al.}(2022){Astropy Collaboration}, {Price-Whelan}, {Lim}, {Earl}, {Starkman}, {Bradley}, {Shupe}, {Patil}, {Corrales}, {Brasseur}, {N{"o}the}, {Donath}, {Tollerud}, {Morris}, {Ginsburg}, {Vaher}, {Weaver}, {Tocknell}, {Jamieson}, {van Kerkwijk}, {Robitaille}, {Merry}, {Bachetti}, {G{"u}nther}, {Aldcroft}, {Alvarado-Montes}, {Archibald}, {B{'o}di}, {Bapat}, {Barentsen}, {Baz{'a}n}, {Biswas}, {Boquien}, {Burke}, {Cara}, {Cara}, {Conroy}, {Conseil}, {Craig}, {Cross}, {Cruz}, {D'Eugenio}, {Dencheva}, {Devillepoix}, {Dietrich}, {Eigenbrot}, {Erben}, {Ferreira}, {Foreman-Mackey}, {Fox}, {Freij}, {Garg}, {Geda}, {Glattly}, {Gondhalekar}, {Gordon}, {Grant}, {Greenfield}, {Groener}, {Guest}, {Gurovich}, {Handberg}, {Hart}, {Hatfield-Dodds}, {Homeier}, {Hosseinzadeh}, {Jenness}, {Jones}, {Joseph}, {Kalmbach}, {Karamehmetoglu}, {Ka{l}uszy{'n}ski}, {Kelley}, {Kern}, {Kerzendorf}, {Koch}, {Kulumani}, {Lee}, {Ly}, {Ma}, {MacBride}, {Maljaars}, {Muna}, {Murphy}, {Norman}, {O'Steen},
  {Oman}, {Pacifici}, {Pascual}, {Pascual-Granado}, {Patil}, {Perren}, {Pickering}, {Rastogi}, {Roulston}, {Ryan}, {Rykoff}, {Sabater}, {Sakurikar}, {Salgado}, {Sanghi}, {Saunders}, {Savchenko}, {Schwardt}, {Seifert-Eckert}, {Shih}, {Jain}, {Shukla}, {Sick}, {Simpson}, {Singanamalla}, {Singer}, {Singhal}, {Sinha}, {Sip{H{o}}cz}, {Spitler}, {Stansby}, {Streicher}, {{{S}}umak}, {Swinbank}, {Taranu}, {Tewary}, {Tremblay}, {Val-Borro}, {Van Kooten}, {Vasovi{'c}}, {Verma}, {de Miranda Cardoso}, {Williams}, {Wilson}, {Winkel}, {Wood-Vasey}, {Xue}, {Yoachim}, {Zhang}, {Zonca}, \& {Astropy Project Contributors}}]{astropy:2022}
{Astropy Collaboration}, {Price-Whelan}, A.~M., {Lim}, P.~L., {et~al.} 2022, \apj, 935, 167, \dodoi{10.3847/1538-4357/ac7c74}

\bibitem[{{Balona}(1992)}]{Balona92}
{Balona}, L.~A. 1992, \mnras, 256, 425, \dodoi{10.1093/mnras/256.3.425}

\bibitem[{{Bj{\"o}rklund} {et~al.}(2021){Bj{\"o}rklund}, {Sundqvist}, {Puls}, \& {Najarro}}]{Bjorklund21}
{Bj{\"o}rklund}, R., {Sundqvist}, J.~O., {Puls}, J., \& {Najarro}, F. 2021, \aap, 648, A36, \dodoi{10.1051/0004-6361/202038384}

\bibitem[{{Bodensteiner} {et~al.}(2021){Bodensteiner}, {Sana}, {Wang}, {Langer}, {Mahy}, {Banyard}, {de Koter}, {de Mink}, {Evans}, {G{\"o}tberg}, {Patrick}, {Schneider}, \& {Tramper}}]{Bodensteiner21}
{Bodensteiner}, J., {Sana}, H., {Wang}, C., {et~al.} 2021, \aap, 652, A70, \dodoi{10.1051/0004-6361/202140507}

\bibitem[{{B{\"o}hm-Vitense}(1958)}]{bohm-vitense1958}
{B{\"o}hm-Vitense}, E. 1958, \zap, 46, 108

\bibitem[{{Bonanos} {et~al.}(2024){Bonanos}, {Tramper}, {de Wit}, {Christodoulou}, {Mu{\~n}oz Sanchez}, {Antoniadis}, {Athanasiou}, {Maravelias}, {Yang}, \& {Zapartas}}]{Bonanos23}
{Bonanos}, A.~Z., {Tramper}, F., {de Wit}, S., {et~al.} 2024, \aap, 686, A77, \dodoi{10.1051/0004-6361/202348527}

\bibitem[{{Bradstreet}(2005)}]{Bradstreet05}
{Bradstreet}, D.~H. 2005, Society for Astronomical Sciences Annual Symposium, 24, 23

\bibitem[{{Briel} {et~al.}(2023){Briel}, {Stevance}, \& {Eldridge}}]{Briel23}
{Briel}, M.~M., {Stevance}, H.~F., \& {Eldridge}, J.~J. 2023, \mnras, 520, 5724, \dodoi{10.1093/mnras/stad399}

\bibitem[{{Buchner}(2016)}]{Buchner16}
{Buchner}, J. 2016, Statistics and Computing, 26, 383, \dodoi{10.1007/s11222-014-9512-y}

\bibitem[{{Buchner}(2019)}]{Buchner19}
---. 2019, \pasp, 131, 108005, \dodoi{10.1088/1538-3873/aae7fc}

\bibitem[{{Buchner}(2021)}]{Buchner21}
---. 2021, The Journal of Open Source Software, 6, 3001, \dodoi{10.21105/joss.03001}

\bibitem[{Buchner(2023)}]{Buchner23}
Buchner, J. 2023, in MaxEnt 2022, MaxEnt 2022 (MDPI), \dodoi{10.3390/psf2022005046}

\bibitem[{{Castelli} \& {Kurucz}(2004)}]{castelli04}
{Castelli}, F., \& {Kurucz}, R.~L. 2004, \aap, 419, 725, \dodoi{10.1051/0004-6361:20040079}

\bibitem[{{Choi} {et~al.}(2016){Choi}, {Dotter}, {Conroy}, {Cantiello}, {Paxton}, \& {Johnson}}]{choi16}
{Choi}, J., {Dotter}, A., {Conroy}, C., {et~al.} 2016, \apj, 823, 102, \dodoi{10.3847/0004-637X/823/2/102}

\bibitem[{{Chu-Qi} {et~al.}(2023){Chu-Qi}, {Fu-Xing}, {Sheng-Bang}, {Jia}, {Sarotsakulchai}, {Zubairi}, \& {Azizbek}}]{Wu23b}
{Chu-Qi}, W., {Fu-Xing}, L., {Sheng-Bang}, Q., {et~al.} 2023, \pasp, 135, 094202, \dodoi{10.1088/1538-3873/acf8f9}

\bibitem[{{Conroy} {et~al.}(2020){Conroy}, {Kochoska}, {Hey}, {Pablo}, {Hambleton}, {Jones}, {Giammarco}, {Abdul-Masih}, \& {Pr{\v{s}}a}}]{Conroy20}
{Conroy}, K.~E., {Kochoska}, A., {Hey}, D., {et~al.} 2020, \apjs, 250, 34, \dodoi{10.3847/1538-4365/abb4e2}

\bibitem[{{Cox} \& {Giuli}(1968)}]{Cox1968}
{Cox}, J.~P., \& {Giuli}, R.~T. 1968, {Principles of stellar structure}

\bibitem[{{Dalcanton} {et~al.}(2009){Dalcanton}, {Williams}, {Seth}, {Dolphin}, {Holtzman}, {Rosema}, {Skillman}, {Cole}, {Girardi}, {Gogarten}, {Karachentsev}, {Olsen}, {Weisz}, {Christensen}, {Freeman}, {Gilbert}, {Gallart}, {Harris}, {Hodge}, {de Jong}, {Karachentseva}, {Mateo}, {Stetson}, {Tavarez}, {Zaritsky}, {Governato}, \& {Quinn}}]{Dalcanton09}
{Dalcanton}, J.~J., {Williams}, B.~F., {Seth}, A.~C., {et~al.} 2009, \apjs, 183, 67, \dodoi{10.1088/0067-0049/183/1/67}

\bibitem[{{de Mink} {et~al.}(2007){de Mink}, {Pols}, \& {Hilditch}}]{demink07}
{de Mink}, S.~E., {Pols}, O.~R., \& {Hilditch}, R.~W. 2007, \aap, 467, 1181, \dodoi{10.1051/0004-6361:20067007}

\bibitem[{{Ding} {et~al.}(2022){Ding}, {Ji}, {Li}, {Xiong}, {Cheng}, {Wang}, \& {Liu}}]{Ding22}
{Ding}, X., {Ji}, K., {Li}, X., {et~al.} 2022, \aj, 164, 200, \dodoi{10.3847/1538-3881/ac8e66}

\bibitem[{{Ding} {et~al.}(2021){Ding}, {Ji}, \& {Li}}]{Ding21}
{Ding}, X., {Ji}, K.-F., \& {Li}, X.-Z. 2021, \pasj, 73, 786, \dodoi{10.1093/pasj/psab042}

\bibitem[{{Dolphin}(2016)}]{DOLPHOT}
{Dolphin}, A. 2016, {DOLPHOT: Stellar photometry}, Astrophysics Source Code Library, record ascl:1608.013.
\newblock \doeprint{1608.013}

\bibitem[{{Dolphin}(2000)}]{Dolphin00}
{Dolphin}, A.~E. 2000, \pasp, 112, 1383, \dodoi{10.1086/316630}

\bibitem[{{Dorozsmai} \& {Toonen}(2024)}]{Dorozsmai24}
{Dorozsmai}, A., \& {Toonen}, S. 2024, \mnras, 530, 3706, \dodoi{10.1093/mnras/stae152}

\bibitem[{{Dotter}(2016)}]{dotter16}
{Dotter}, A. 2016, \apjs, 222, 8, \dodoi{10.3847/0067-0049/222/1/8}

\bibitem[{{Dufton} {et~al.}(2019){Dufton}, {Evans}, {Hunter}, {Lennon}, \& {Schneider}}]{Dufton19}
{Dufton}, P.~L., {Evans}, C.~J., {Hunter}, I., {Lennon}, D.~J., \& {Schneider}, F.~R.~N. 2019, \aap, 626, A50, \dodoi{10.1051/0004-6361/201935415}

\bibitem[{{Dunstall} {et~al.}(2015){Dunstall}, {Dufton}, {Sana}, {Evans}, {Howarth}, {Sim{\'o}n-D{\'\i}az}, {de Mink}, {Langer}, {Ma{\'\i}z Apell{\'a}niz}, \& {Taylor}}]{Dunstall15}
{Dunstall}, P.~R., {Dufton}, P.~L., {Sana}, H., {et~al.} 2015, \aap, 580, A93, \dodoi{10.1051/0004-6361/201526192}

\bibitem[{{Eldridge} {et~al.}(2013){Eldridge}, {Fraser}, {Smartt}, {Maund}, \& {Crockett}}]{Eldridge13}
{Eldridge}, J.~J., {Fraser}, M., {Smartt}, S.~J., {Maund}, J.~R., \& {Crockett}, R.~M. 2013, \mnras, 436, 774, \dodoi{10.1093/mnras/stt1612}

\bibitem[{{Eldridge} \& {Stanway}(2020)}]{Eldridge20}
{Eldridge}, J.~J., \& {Stanway}, E.~R. 2020, arXiv e-prints, arXiv:2005.11883, \dodoi{10.48550/arXiv.2005.11883}

\bibitem[{{Eldridge} \& {Stanway}(2022)}]{eldridge22}
---. 2022, \araa, 60, 455, \dodoi{10.1146/annurev-astro-052920-100646}

\bibitem[{{Evans} {et~al.}(2023){Evans}, {Marcolino}, {Bouret}, \& {Garcia}}]{Evans23}
{Evans}, C., {Marcolino}, W., {Bouret}, J.-C., \& {Garcia}, M. 2023, Experimental Astronomy, 56, 537, \dodoi{10.1007/s10686-023-09912-w}

\bibitem[{{Fabry} {et~al.}(2023){Fabry}, {Marchant}, {Langer}, \& {Sana}}]{Fabry23}
{Fabry}, M., {Marchant}, P., {Langer}, N., \& {Sana}, H. 2023, \aap, 672, A175, \dodoi{10.1051/0004-6361/202346277}

\bibitem[{{Fabry} {et~al.}(2024){Fabry}, {Marchant}, {Langer}, \& {Sana}}]{Fabry24}
---. 2024, arXiv e-prints, arXiv:2410.21394, \dodoi{10.48550/arXiv.2410.21394}

\bibitem[{{Fabry} {et~al.}(2022){Fabry}, {Marchant}, \& {Sana}}]{Fabry22}
{Fabry}, M., {Marchant}, P., \& {Sana}, H. 2022, \aap, 661, A123, \dodoi{10.1051/0004-6361/202243094}

\bibitem[{Foreman-Mackey(2016)}]{corner}
Foreman-Mackey, D. 2016, The Journal of Open Source Software, 1, 24, \dodoi{10.21105/joss.00024}

\bibitem[{{Gilbert} {et~al.}(2025){Gilbert}, {Choi}, {Boyer}, {Williams}, {Weisz}, {Bell}, {Dalcanton}, {McQuinn}, {Skillman}, {Costa}, {Dolphin}, {Fouesneau}, {Girardi}, {Goldman}, {Gordon}, {Guhathakurta}, {Gull}, {Hagen}, {Huynh}, {Lindberg}, {Marigo}, {Murray}, {Pastorelli}, \& {Yanchulova Merica-Jones}}]{Gilbert25}
{Gilbert}, K.~M., {Choi}, Y., {Boyer}, M.~L., {et~al.} 2025, \apjs, 276, 8, \dodoi{10.3847/1538-4365/ad76af}

\bibitem[{{Goldman} {et~al.}(2019){Goldman}, {Boyer}, {McQuinn}, {Whitelock}, {McDonald}, {van Loon}, {Skillman}, {Gehrz}, {Javadi}, {Sloan}, {Jones}, {Groenewegen}, \& {Menzies}}]{Goldman19}
{Goldman}, S.~R., {Boyer}, M.~L., {McQuinn}, K.~B.~W., {et~al.} 2019, \apj, 877, 49, \dodoi{10.3847/1538-4357/ab0965}

\bibitem[{{Gonz{\'a}lez-Tor{\`a}} {et~al.}(2021){Gonz{\'a}lez-Tor{\`a}}, {Davies}, {Kudritzki}, \& {Plez}}]{Gonzalestora21}
{Gonz{\'a}lez-Tor{\`a}}, G., {Davies}, B., {Kudritzki}, R.-P., \& {Plez}, B. 2021, \mnras, 505, 4422, \dodoi{10.1093/mnras/stab1611}

\bibitem[{{G{\"o}tberg} {et~al.}(2017){G{\"o}tberg}, {de Mink}, \& {Groh}}]{Gotberg17}
{G{\"o}tberg}, Y., {de Mink}, S.~E., \& {Groh}, J.~H. 2017, \aap, 608, A11, \dodoi{10.1051/0004-6361/201730472}

\bibitem[{{Groot} {et~al.}(2022){Groot}, {Bloemen}, {Vreeswijk}, {Jonker}, {Pieterse}, {Engels}, {Michiels}, {Bakker}, {Hahn}, {Raskin}, {Morren}, {Navarro}, {Elswijk}, {ter Horst}, {Schuil}, {Kragt}, {Lesman}, {de Haan}, {Bekema}, {de Haan}, {Klein-Wolt}, {Blagorodnova}, {Johnston}, \& {Le Poole}}]{Groot22}
{Groot}, P.~J., {Bloemen}, S., {Vreeswijk}, P.~M., {et~al.} 2022, in Society of Photo-Optical Instrumentation Engineers (SPIE) Conference Series, Vol. 12182, Ground-based and Airborne Telescopes IX, ed. H.~K. {Marshall}, J.~{Spyromilio}, \& T.~{Usuda}, 121821V, \dodoi{10.1117/12.2630160}

\bibitem[{{Gull} {et~al.}(2022){Gull}, {Weisz}, {Senchyna}, {Sandford}, {Choi}, {McLeod}, {El-Badry}, {G{\"o}tberg}, {Gilbert}, {Boyer}, {Dalcanton}, {GuhaThakurta}, {Goldman}, {Marigo}, {McQuinn}, {Pastorelli}, {Stark}, {Skillman}, {Ting}, \& {Williams}}]{Gull22}
{Gull}, M., {Weisz}, D.~R., {Senchyna}, P., {et~al.} 2022, \apj, 941, 206, \dodoi{10.3847/1538-4357/aca295}

\bibitem[{{Hambleton} {et~al.}(2023){Hambleton}, {Bianco}, {Street}, {Bell}, {Buckley}, {Graham}, {Hernitschek}, {Lund}, {Mason}, {Pepper}, {Pr{\v{s}}a}, {Rabus}, {Raiteri}, {Szab{\'o}}, {Szkody}, {Andreoni}, {Antoniucci}, {Balmaverde}, {Bellm}, {Bonito}, {Bono}, {Botticella}, {Brocato}, {Bu{\v{c}}ar Bricman}, {Cappellaro}, {Carnerero}, {Chornock}, {Clarke}, {Cowperthwaite}, {Cucchiara}, {D'Ammando}, {Dage}, {Dall'Ora}, {Davenport}, {de Martino}, {de Somma}, {Di Criscienzo}, {Di Stefano}, {Drout}, {Fabrizio}, {Fiorentino}, {Gandhi}, {Garofalo}, {Giannini}, {Gomboc}, {Greggio}, {Hartigan}, {Hundertmark}, {Johnson}, {Johnson}, {Jurkic}, {Khakpash}, {Leccia}, {Li}, {Magurno}, {Malanchev}, {Marconi}, {Margutti}, {Marinoni}, {Mauron}, {Molinaro}, {M{\"o}ller}, {Moniez}, {Muraveva}, {Musella}, {Ngeow}, {Pastorello}, {Petrecca}, {Piranomonte}, {Ragosta}, {Reguitti}, {Righi}, {Ripepi}, {Rivera Sandoval}, {Stassun}, {Stroh}, {Terreran}, {Trimble}, {Tsapras}, {van Velzen}, {Venuti}, \& {Vink}}]{Hambleton23}
{Hambleton}, K.~M., {Bianco}, F.~B., {Street}, R., {et~al.} 2023, \pasp, 135, 105002, \dodoi{10.1088/1538-3873/acdb9a}

\bibitem[{Harris {et~al.}(2020)Harris, Millman, van~der Walt, Gommers, Virtanen, Cournapeau, Wieser, Taylor, Berg, Smith, Kern, Picus, Hoyer, van Kerkwijk, Brett, Haldane, del R{\'{i}}o, Wiebe, Peterson, G{\'{e}}rard-Marchant, Sheppard, Reddy, Weckesser, Abbasi, Gohlke, \& Oliphant}]{numpy}
Harris, C.~R., Millman, K.~J., van~der Walt, S.~J., {et~al.} 2020, Nature, 585, 357, \dodoi{10.1038/s41586-020-2649-2}

\bibitem[{{Hastings} {et~al.}(2020){Hastings}, {Langer}, \& {Koenigsberger}}]{Hastings20}
{Hastings}, B., {Langer}, N., \& {Koenigsberger}, G. 2020, \aap, 641, A86, \dodoi{10.1051/0004-6361/202038499}

\bibitem[{{Henneco} {et~al.}(2024){Henneco}, {Schneider}, \& {Laplace}}]{Henneco2024}
{Henneco}, J., {Schneider}, F.~R.~N., \& {Laplace}, E. 2024, \aap, 682, A169, \dodoi{10.1051/0004-6361/202347893}

\bibitem[{{Hilditch} {et~al.}(2005){Hilditch}, {Howarth}, \& {Harries}}]{Hilditch05}
{Hilditch}, R.~W., {Howarth}, I.~D., \& {Harries}, T.~J. 2005, \mnras, 357, 304, \dodoi{10.1111/j.1365-2966.2005.08653.x}

\bibitem[{{Horvat} {et~al.}(2018){Horvat}, {Conroy}, {Pablo}, {Hambleton}, {Kochoska}, {Giammarco}, \& {Pr{\v{s}}a}}]{Horvat18}
{Horvat}, M., {Conroy}, K.~E., {Pablo}, H., {et~al.} 2018, \apjs, 237, 26, \dodoi{10.3847/1538-4365/aacd0f}

\bibitem[{Hunter(2007)}]{matplotlib}
Hunter, J.~D. 2007, Computing in Science \& Engineering, 9, 90, \dodoi{10.1109/MCSE.2007.55}

\bibitem[{{Hurley} {et~al.}(2002){Hurley}, {Tout}, \& {Pols}}]{hurley02}
{Hurley}, J.~R., {Tout}, C.~A., \& {Pols}, O.~R. 2002, \mnras, 329, 897, \dodoi{10.1046/j.1365-8711.2002.05038.x}

\bibitem[{{Ivezi{\'c}} {et~al.}(2019){Ivezi{\'c}}, {Kahn}, {Tyson}, {Abel}, {Acosta}, {Allsman}, {Alonso}, {AlSayyad}, {Anderson}, {Andrew}, {Angel}, {Angeli}, {Ansari}, {Antilogus}, {Araujo}, {Armstrong}, {Arndt}, {Astier}, {Aubourg}, {Auza}, {Axelrod}, {Bard}, {Barr}, {Barrau}, {Bartlett}, {Bauer}, {Bauman}, {Baumont}, {Bechtol}, {Bechtol}, {Becker}, {Becla}, {Beldica}, {Bellavia}, {Bianco}, {Biswas}, {Blanc}, {Blazek}, {Blandford}, {Bloom}, {Bogart}, {Bond}, {Booth}, {Borgland}, {Borne}, {Bosch}, {Boutigny}, {Brackett}, {Bradshaw}, {Brandt}, {Brown}, {Bullock}, {Burchat}, {Burke}, {Cagnoli}, {Calabrese}, {Callahan}, {Callen}, {Carlin}, {Carlson}, {Chandrasekharan}, {Charles-Emerson}, {Chesley}, {Cheu}, {Chiang}, {Chiang}, {Chirino}, {Chow}, {Ciardi}, {Claver}, {Cohen-Tanugi}, {Cockrum}, {Coles}, {Connolly}, {Cook}, {Cooray}, {Covey}, {Cribbs}, {Cui}, {Cutri}, {Daly}, {Daniel}, {Daruich}, {Daubard}, {Daues}, {Dawson}, {Delgado}, {Dellapenna}, {de Peyster}, {de Val-Borro}, {Digel}, {Doherty}, {Dubois},
  {Dubois-Felsmann}, {Durech}, {Economou}, {Eifler}, {Eracleous}, {Emmons}, {Fausti Neto}, {Ferguson}, {Figueroa}, {Fisher-Levine}, {Focke}, {Foss}, {Frank}, {Freemon}, {Gangler}, {Gawiser}, {Geary}, {Gee}, {Geha}, {Gessner}, {Gibson}, {Gilmore}, {Glanzman}, {Glick}, {Goldina}, {Goldstein}, {Goodenow}, {Graham}, {Gressler}, {Gris}, {Guy}, {Guyonnet}, {Haller}, {Harris}, {Hascall}, {Haupt}, {Hernandez}, {Herrmann}, {Hileman}, {Hoblitt}, {Hodgson}, {Hogan}, {Howard}, {Huang}, {Huffer}, {Ingraham}, {Innes}, {Jacoby}, {Jain}, {Jammes}, {Jee}, {Jenness}, {Jernigan}, {Jevremovi{\'c}}, {Johns}, {Johnson}, {Johnson}, {Jones}, {Juramy-Gilles}, {Juri{\'c}}, {Kalirai}, {Kallivayalil}, {Kalmbach}, {Kantor}, {Karst}, {Kasliwal}, {Kelly}, {Kessler}, {Kinnison}, {Kirkby}, {Knox}, {Kotov}, {Krabbendam}, {Krughoff}, {Kub{\'a}nek}, {Kuczewski}, {Kulkarni}, {Ku}, {Kurita}, {Lage}, {Lambert}, {Lange}, {Langton}, {Le Guillou}, {Levine}, {Liang}, {Lim}, {Lintott}, {Long}, {Lopez}, {Lotz}, {Lupton}, {Lust}, {MacArthur}, {Mahabal},
  {Mandelbaum}, {Markiewicz}, {Marsh}, {Marshall}, {Marshall}, {May}, {McKercher}, {McQueen}, {Meyers}, {Migliore}, {Miller}, {Mills}, {Miraval}, {Moeyens}, {Moolekamp}, {Monet}, {Moniez}, {Monkewitz}, {Montgomery}, {Morrison}, {Mueller}, {Muller}, {Mu{\~n}oz Arancibia}, {Neill}, {Newbry}, {Nief}, {Nomerotski}, {Nordby}, {O'Connor}, {Oliver}, {Olivier}, {Olsen}, {O'Mullane}, {Ortiz}, {Osier}, {Owen}, {Pain}, {Palecek}, {Parejko}, {Parsons}, {Pease}, {Peterson}, {Peterson}, {Petravick}, {Libby Petrick}, {Petry}, {Pierfederici}, {Pietrowicz}, {Pike}, {Pinto}, {Plante}, {Plate}, {Plutchak}, {Price}, {Prouza}, {Radeka}, {Rajagopal}, {Rasmussen}, {Regnault}, {Reil}, {Reiss}, {Reuter}, {Ridgway}, {Riot}, {Ritz}, {Robinson}, {Roby}, {Roodman}, {Rosing}, {Roucelle}, {Rumore}, {Russo}, {Saha}, {Sassolas}, {Schalk}, {Schellart}, {Schindler}, {Schmidt}, {Schneider}, {Schneider}, {Schoening}, {Schumacher}, {Schwamb}, {Sebag}, {Selvy}, {Sembroski}, {Seppala}, {Serio}, {Serrano}, {Shaw}, {Shipsey}, {Sick}, {Silvestri},
  {Slater}, {Smith}, {Smith}, {Sobhani}, {Soldahl}, {Storrie-Lombardi}, {Stover}, {Strauss}, {Street}, {Stubbs}, {Sullivan}, {Sweeney}, {Swinbank}, {Szalay}, {Takacs}, {Tether}, {Thaler}, {Thayer}, {Thomas}, {Thornton}, {Thukral}, {Tice}, {Trilling}, {Turri}, {Van Berg}, {Vanden Berk}, {Vetter}, {Virieux}, {Vucina}, {Wahl}, {Walkowicz}, {Walsh}, {Walter}, {Wang}, {Wang}, {Warner}, {Wiecha}, {Willman}, {Winters}, {Wittman}, {Wolff}, {Wood-Vasey}, {Wu}, {Xin}, {Yoachim}, \& {Zhan}}]{Ivezic19}
{Ivezi{\'c}}, {\v{Z}}., {Kahn}, S.~M., {Tyson}, J.~A., {et~al.} 2019, \apj, 873, 111, \dodoi{10.3847/1538-4357/ab042c}

\bibitem[{{Jones} {et~al.}(1973){Jones}, {Forman}, {Tananbaum}, {Schreier}, {Gursky}, {Kellogg}, \& {Giacconi}}]{Jones73}
{Jones}, C., {Forman}, W., {Tananbaum}, H., {et~al.} 1973, \apjl, 181, L43, \dodoi{10.1086/181181}

\bibitem[{{Jones} {et~al.}(2020){Jones}, {Conroy}, {Horvat}, {Giammarco}, {Kochoska}, {Pablo}, {Brown}, {Sowicka}, \& {Pr{\v{s}}a}}]{Jones20}
{Jones}, D., {Conroy}, K.~E., {Horvat}, M., {et~al.} 2020, \apjs, 247, 63, \dodoi{10.3847/1538-4365/ab7927}

\bibitem[{{Karachentsev} {et~al.}(2013){Karachentsev}, {Makarov}, \& {Kaisina}}]{Karachentsev13}
{Karachentsev}, I.~D., {Makarov}, D.~I., \& {Kaisina}, E.~I. 2013, \aj, 145, 101, \dodoi{10.1088/0004-6256/145/4/101}

\bibitem[{{Kasen} {et~al.}(2017){Kasen}, {Metzger}, {Barnes}, {Quataert}, \& {Ramirez-Ruiz}}]{Kasen17}
{Kasen}, D., {Metzger}, B., {Barnes}, J., {Quataert}, E., \& {Ramirez-Ruiz}, E. 2017, \nat, 551, 80, \dodoi{10.1038/nature24453}

\bibitem[{Kingma \& Ba(2017)}]{kingma17}
Kingma, D.~P., \& Ba, J. 2017, Adam: A Method for Stochastic Optimization.
\newblock \doarXiv{1412.6980}

\bibitem[{{Klencki} {et~al.}(2018){Klencki}, {Moe}, {Gladysz}, {Chruslinska}, {Holz}, \& {Belczynski}}]{Klencki18}
{Klencki}, J., {Moe}, M., {Gladysz}, W., {et~al.} 2018, \aap, 619, A77, \dodoi{10.1051/0004-6361/201833025}

\bibitem[{{Klencki} {et~al.}(2020){Klencki}, {Nelemans}, {Istrate}, \& {Pols}}]{Klencki20}
{Klencki}, J., {Nelemans}, G., {Istrate}, A.~G., \& {Pols}, O. 2020, \aap, 638, A55, \dodoi{10.1051/0004-6361/202037694}

\bibitem[{{Kobulnicky} {et~al.}(2014){Kobulnicky}, {Kiminki}, {Lundquist}, {Burke}, {Chapman}, {Keller}, {Lester}, {Rolen}, {Topel}, {Bhattacharjee}, {Smullen}, {Vargas {\'A}lvarez}, {Runnoe}, {Dale}, \& {Brotherton}}]{Kobulnicky14}
{Kobulnicky}, H.~A., {Kiminki}, D.~C., {Lundquist}, M.~J., {et~al.} 2014, \apjs, 213, 34, \dodoi{10.1088/0067-0049/213/2/34}

\bibitem[{{Kulkarni} {et~al.}(2021){Kulkarni}, {Harrison}, {Grefenstette}, {Earnshaw}, {Andreoni}, {Berg}, {Bloom}, {Cenko}, {Chornock}, {Christiansen}, {Coughlin}, {Wuollet Criswell}, {Darvish}, {Das}, {De}, {Dessart}, {Dixon}, {Dorsman}, {El-Badry}, {Evans}, {Ford}, {Fremling}, {Gansicke}, {Gezari}, {Goetberg}, {Green}, {Graham}, {Heida}, {Ho}, {Jaodand}, {Johns-Krull}, {Kasliwal}, {Lazzarini}, {Lu}, {Margutti}, {Martin}, {Masters}, {McKernan}, {Naze}, {Nissanke}, {Parazin}, {Perley}, {Phinney}, {Piro}, {Raaijmakers}, {Rauw}, {Rodriguez}, {Sana}, {Senchyna}, {Singer}, {Spake}, {Stassun}, {Stern}, {Teplitz}, {Weisz}, \& {Yao}}]{Kulkarni21}
{Kulkarni}, S.~R., {Harrison}, F.~A., {Grefenstette}, B.~W., {et~al.} 2021, arXiv e-prints, arXiv:2111.15608, \dodoi{10.48550/arXiv.2111.15608}

\bibitem[{{Lamb} {et~al.}(2016){Lamb}, {Oey}, {Segura-Cox}, {Graus}, {Kiminki}, {Golden-Marx}, \& {Parker}}]{lamb16}
{Lamb}, J.~B., {Oey}, M.~S., {Segura-Cox}, D.~M., {et~al.} 2016, \apj, 817, 113, \dodoi{10.3847/0004-637X/817/2/113}

\bibitem[{{Langer}(2012)}]{Langer12}
{Langer}, N. 2012, \araa, 50, 107, \dodoi{10.1146/annurev-astro-081811-125534}

\bibitem[{{Laplace} {et~al.}(2021){Laplace}, {Justham}, {Renzo}, {G{\"o}tberg}, {Farmer}, {Vartanyan}, \& {de Mink}}]{Laplace21}
{Laplace}, E., {Justham}, S., {Renzo}, M., {et~al.} 2021, \aap, 656, A58, \dodoi{10.1051/0004-6361/202140506}

\bibitem[{{Lattimer} \& {Schramm}(1976)}]{Lattimer76}
{Lattimer}, J.~M., \& {Schramm}, D.~N. 1976, \apj, 210, 549, \dodoi{10.1086/154860}

\bibitem[{{Lee} {et~al.}(2021){Lee}, {Freedman}, {Madore}, {Owens}, {Monson}, \& {Hoyt}}]{Lee21}
{Lee}, A.~J., {Freedman}, W.~L., {Madore}, B.~F., {et~al.} 2021, \apj, 907, 112, \dodoi{10.3847/1538-4357/abd253}

\bibitem[{{Lee} {et~al.}(2005){Lee}, {Skillman}, \& {Venn}}]{Lee05}
{Lee}, H., {Skillman}, E.~D., \& {Venn}, K.~A. 2005, \apj, 620, 223, \dodoi{10.1086/427019}

\bibitem[{{Lorenzo} {et~al.}(2014){Lorenzo}, {Negueruela}, {Baker}, {Garc{\'\i}a}, {Sim{\'o}n-D{\'\i}az}, {Pastor}, \& {M{\'e}ndez Majuelos}}]{lorenzo14}
{Lorenzo}, J., {Negueruela}, I., {Baker}, A.~K.~F.~V., {et~al.} 2014, \aap, 572, A110, \dodoi{10.1051/0004-6361/201424345}

\bibitem[{{Mahy} {et~al.}(2020){Mahy}, {Almeida}, {Sana}, {Clark}, {de Koter}, {de Mink}, {Evans}, {Grin}, {Langer}, {Moffat}, {Schneider}, {Shenar}, \& {Tramper}}]{Mahy20}
{Mahy}, L., {Almeida}, L.~A., {Sana}, H., {et~al.} 2020, \aap, 634, A119, \dodoi{10.1051/0004-6361/201936152}

\bibitem[{{Mahy} {et~al.}(2022){Mahy}, {Lanthermann}, {Hutsem{\'e}kers}, {Kluska}, {Lobel}, {Manick}, {Miszalski}, {Reggiani}, {Sana}, \& {Gosset}}]{Mahy22}
{Mahy}, L., {Lanthermann}, C., {Hutsem{\'e}kers}, D., {et~al.} 2022, \aap, 657, A4, \dodoi{10.1051/0004-6361/202040062}

\bibitem[{{Mandel} \& {Farmer}(2022)}]{Mandel22}
{Mandel}, I., \& {Farmer}, A. 2022, \physrep, 955, 1, \dodoi{10.1016/j.physrep.2022.01.003}

\bibitem[{{Marchant} \& {Bodensteiner}(2024)}]{Marchant23}
{Marchant}, P., \& {Bodensteiner}, J. 2024, \araa, 62, 21, \dodoi{10.1146/annurev-astro-052722-105936}

\bibitem[{{Marchant} {et~al.}(2016){Marchant}, {Langer}, {Podsiadlowski}, {Tauris}, \& {Moriya}}]{Marchant16}
{Marchant}, P., {Langer}, N., {Podsiadlowski}, P., {Tauris}, T.~M., \& {Moriya}, T.~J. 2016, \aap, 588, A50, \dodoi{10.1051/0004-6361/201628133}

\bibitem[{{Marchant} {et~al.}(2021){Marchant}, {Pappas}, {Gallegos-Garcia}, {Berry}, {Taam}, {Kalogera}, \& {Podsiadlowski}}]{marchant2021}
{Marchant}, P., {Pappas}, K. M.~W., {Gallegos-Garcia}, M., {et~al.} 2021, \aap, 650, A107, \dodoi{10.1051/0004-6361/202039992}

\bibitem[{{Margutti} \& {Chornock}(2021)}]{Margutti21}
{Margutti}, R., \& {Chornock}, R. 2021, \araa, 59, 155, \dodoi{10.1146/annurev-astro-112420-030742}

\bibitem[{{Martins} {et~al.}(2017){Martins}, {Mahy}, \& {Herv{\'e}}}]{Martins17}
{Martins}, F., {Mahy}, L., \& {Herv{\'e}}, A. 2017, \aap, 607, A82, \dodoi{10.1051/0004-6361/201731593}

\bibitem[{{Massey} {et~al.}(2007){Massey}, {McNeill}, {Olsen}, {Hodge}, {Blaha}, {Jacoby}, {Smith}, \& {Strong}}]{massey07}
{Massey}, P., {McNeill}, R.~T., {Olsen}, K.~A.~G., {et~al.} 2007, \aj, 134, 2474, \dodoi{10.1086/523658}

\bibitem[{{McQuinn} {et~al.}(2024){McQuinn}, {B. Newman}, {Savino}, {Dolphin}, {Weisz}, {Williams}, {Boyer}, {Cohen}, {Correnti}, {Cole}, {Geha}, {Gennaro}, {Kallivayalil}, {Sandstrom}, {Skillman}, {Anderson}, {Bolatto}, {Boylan-Kolchin}, {Garling}, {Gilbert}, {Girardi}, {Kalirai}, {Mazzi}, {Pastorelli}, {Richstein}, \& {Warfield}}]{McQuinn24}
{McQuinn}, K. B.~W., {B. Newman}, M.~J., {Savino}, A., {et~al.} 2024, \apj, 961, 16, \dodoi{10.3847/1538-4357/ad1105}

\bibitem[{{Menon} {et~al.}(2024){Menon}, {Pawlak}, {Lennon}, {Sen}, \& {Langer}}]{Menon24}
{Menon}, A., {Pawlak}, M., {Lennon}, D.~J., {Sen}, K., \& {Langer}, N. 2024, arXiv e-prints, arXiv:2410.16427, \dodoi{10.48550/arXiv.2410.16427}

\bibitem[{{Menon} {et~al.}(2021){Menon}, {Langer}, {de Mink}, {Justham}, {Sen}, {Sz{\'e}csi}, {de Koter}, {Abdul-Masih}, {Sana}, {Mahy}, \& {Marchant}}]{Menon21}
{Menon}, A., {Langer}, N., {de Mink}, S.~E., {et~al.} 2021, \mnras, 507, 5013, \dodoi{10.1093/mnras/stab2276}

\bibitem[{{Moe} \& {Di Stefano}(2013)}]{Moe13}
{Moe}, M., \& {Di Stefano}, R. 2013, \apj, 778, 95, \dodoi{10.1088/0004-637X/778/2/95}

\bibitem[{{Moe} \& {Di Stefano}(2015)}]{moe15}
---. 2015, \apj, 810, 61, \dodoi{10.1088/0004-637X/810/1/61}

\bibitem[{{Moe} \& {Di Stefano}(2017)}]{Moe17}
---. 2017, \apjs, 230, 15, \dodoi{10.3847/1538-4365/aa6fb6}

\bibitem[{{Muraveva} {et~al.}(2014){Muraveva}, {Clementini}, {Maceroni}, {Evans}, {Moretti}, {Cioni}, {Marquette}, {Ripepi}, {de Grijs}, {Groenewegen}, {Piatti}, \& {van Loon}}]{Muraveva14}
{Muraveva}, T., {Clementini}, G., {Maceroni}, C., {et~al.} 2014, \mnras, 443, 432, \dodoi{10.1093/mnras/stu1151}

\bibitem[{{North} {et~al.}(2010){North}, {Gauderon}, {Barblan}, \& {Royer}}]{North10}
{North}, P., {Gauderon}, R., {Barblan}, F., \& {Royer}, F. 2010, \aap, 520, A74, \dodoi{10.1051/0004-6361/200810284}

\bibitem[{{Offner} {et~al.}(2023){Offner}, {Moe}, {Kratter}, {Sadavoy}, {Jensen}, \& {Tobin}}]{Offner23}
{Offner}, S.~S.~R., {Moe}, M., {Kratter}, K.~M., {et~al.} 2023, in Astronomical Society of the Pacific Conference Series, Vol. 534, Protostars and Planets VII, ed. S.~{Inutsuka}, Y.~{Aikawa}, T.~{Muto}, K.~{Tomida}, \& M.~{Tamura}, 275, \dodoi{10.48550/arXiv.2203.10066}

\bibitem[{{Paczy{\'n}ski}(1967)}]{Pacznski67}
{Paczy{\'n}ski}, B. 1967, \actaa, 17, 355

\bibitem[{Paszke {et~al.}(2019)Paszke, Gross, Massa, Lerer, Bradbury, Chanan, Killeen, Lin, Gimelshein, Antiga, Desmaison, Köpf, Yang, DeVito, Raison, Tejani, Chilamkurthy, Steiner, Fang, Bai, \& Chintala}]{pytorch}
Paszke, A., Gross, S., Massa, F., {et~al.} 2019, PyTorch: An Imperative Style, High-Performance Deep Learning Library.
\newblock \doarXiv{1912.01703}

\bibitem[{{Pawlak}(2016)}]{Pawlak16}
{Pawlak}, M. 2016, \mnras, 457, 4323, \dodoi{10.1093/mnras/stw269}

\bibitem[{{Paxton} {et~al.}(2011){Paxton}, {Bildsten}, {Dotter}, {Herwig}, {Lesaffre}, \& {Timmes}}]{Paxton11}
{Paxton}, B., {Bildsten}, L., {Dotter}, A., {et~al.} 2011, \apjs, 192, 3, \dodoi{10.1088/0067-0049/192/1/3}

\bibitem[{{Paxton} {et~al.}(2013){Paxton}, {Cantiello}, {Arras}, {Bildsten}, {Brown}, {Dotter}, {Mankovich}, {Montgomery}, {Stello}, {Timmes}, \& {Townsend}}]{Paxton13}
{Paxton}, B., {Cantiello}, M., {Arras}, P., {et~al.} 2013, \apjs, 208, 4, \dodoi{10.1088/0067-0049/208/1/4}

\bibitem[{{Paxton} {et~al.}(2015){Paxton}, {Marchant}, {Schwab}, {Bauer}, {Bildsten}, {Cantiello}, {Dessart}, {Farmer}, {Hu}, {Langer}, {Townsend}, {Townsley}, \& {Timmes}}]{Paxton15}
{Paxton}, B., {Marchant}, P., {Schwab}, J., {et~al.} 2015, \apjs, 220, 15, \dodoi{10.1088/0067-0049/220/1/15}

\bibitem[{{Paxton} {et~al.}(2018){Paxton}, {Schwab}, {Bauer}, {Bildsten}, {Blinnikov}, {Duffell}, {Farmer}, {Goldberg}, {Marchant}, {Sorokina}, {Thoul}, {Townsend}, \& {Timmes}}]{Paxton18}
{Paxton}, B., {Schwab}, J., {Bauer}, E.~B., {et~al.} 2018, \apjs, 234, 34, \dodoi{10.3847/1538-4365/aaa5a8}

\bibitem[{{Paxton} {et~al.}(2019){Paxton}, {Smolec}, {Schwab}, {Gautschy}, {Bildsten}, {Cantiello}, {Dotter}, {Farmer}, {Goldberg}, {Jermyn}, {Kanbur}, {Marchant}, {Thoul}, {Townsend}, {Wolf}, {Zhang}, \& {Timmes}}]{Paxton19}
{Paxton}, B., {Smolec}, R., {Schwab}, J., {et~al.} 2019, \apjs, 243, 10, \dodoi{10.3847/1538-4365/ab2241}

\bibitem[{{Penny} {et~al.}(2008){Penny}, {Ouzts}, \& {Gies}}]{Penny08}
{Penny}, L.~R., {Ouzts}, C., \& {Gies}, D.~R. 2008, \apj, 681, 554, \dodoi{10.1086/587509}

\bibitem[{{Podsiadlowski} {et~al.}(1992){Podsiadlowski}, {Joss}, \& {Hsu}}]{Podsiadlowski92}
{Podsiadlowski}, P., {Joss}, P.~C., \& {Hsu}, J.~J.~L. 1992, \apj, 391, 246, \dodoi{10.1086/171341}

\bibitem[{{Pols}(1994)}]{Pols94}
{Pols}, O.~R. 1994, \aap, 290, 119

\bibitem[{{Pr{\v{s}}a} \& {Zwitter}(2005)}]{Prsa05}
{Pr{\v{s}}a}, A., \& {Zwitter}, T. 2005, \apj, 628, 426, \dodoi{10.1086/430591}

\bibitem[{{Pr{\v{s}}a} {et~al.}(2016){Pr{\v{s}}a}, {Conroy}, {Horvat}, {Pablo}, {Kochoska}, {Bloemen}, {Giammarco}, {Hambleton}, \& {Degroote}}]{Prsa16}
{Pr{\v{s}}a}, A., {Conroy}, K.~E., {Horvat}, M., {et~al.} 2016, \apjs, 227, 29, \dodoi{10.3847/1538-4365/227/2/29}

\bibitem[{{Ramachandran} {et~al.}(2024){Ramachandran}, {Sander}, {Pauli}, {Klencki}, {Backs}, {Tramper}, {Bernini-Peron}, {Crowther}, {Hamann}, {Ignace}, {Kuiper}, {Oey}, {Oskinova}, {Shenar}, {Todt}, {Vink}, {Wang}, {Wofford}, \& {the XShootU Collaboration}}]{Ramachandran24}
{Ramachandran}, V., {Sander}, A.~A.~C., {Pauli}, D., {et~al.} 2024, \aap, 692, A90, \dodoi{10.1051/0004-6361/202449665}

\bibitem[{{Raucq} {et~al.}(2017){Raucq}, {Gosset}, {Rauw}, {Manfroid}, {Mahy}, {Mennekens}, \& {Vanbeveren}}]{Rauc17}
{Raucq}, F., {Gosset}, E., {Rauw}, G., {et~al.} 2017, \aap, 601, A133, \dodoi{10.1051/0004-6361/201630330}

\bibitem[{{Rickard} \& {Pauli}(2023)}]{Rickard23}
{Rickard}, M.~J., \& {Pauli}, D. 2023, \aap, 674, A56, \dodoi{10.1051/0004-6361/202346055}

\bibitem[{{Rizzuto} {et~al.}(2013){Rizzuto}, {Ireland}, {Robertson}, {Kok}, {Tuthill}, {Warrington}, {Haubois}, {Tango}, {Norris}, {ten Brummelaar}, {Kraus}, {Jacob}, \& {Laliberte-Houdeville}}]{Rizzuto13}
{Rizzuto}, A.~C., {Ireland}, M.~J., {Robertson}, J.~G., {et~al.} 2013, \mnras, 436, 1694, \dodoi{10.1093/mnras/stt1690}

\bibitem[{{Sana} {et~al.}(2013){Sana}, {de Koter}, {de Mink}, {Dunstall}, {Evans}, {H{\'e}nault-Brunet}, {Ma{\'\i}z Apell{\'a}niz}, {Ram{\'\i}rez-Agudelo}, {Taylor}, {Walborn}, {Clark}, {Crowther}, {Herrero}, {Gieles}, {Langer}, {Lennon}, \& {Vink}}]{Sana2013}
{Sana}, H., {de Koter}, A., {de Mink}, S.~E., {et~al.} 2013, \aap, 550, A107, \dodoi{10.1051/0004-6361/201219621}

\bibitem[{{Savino} {et~al.}(2022){Savino}, {Weisz}, {Skillman}, {Dolphin}, {Kallivayalil}, {Wetzel}, {Anderson}, {Besla}, {Boylan-Kolchin}, {Bullock}, {Cole}, {Collins}, {Cooper}, {Deason}, {Dotter}, {Fardal}, {Ferguson}, {Fritz}, {Geha}, {Gilbert}, {Guhathakurta}, {Ibata}, {Irwin}, {Jeon}, {Kirby}, {Lewis}, {Mackey}, {Majewski}, {Martin}, {McConnachie}, {Patel}, {Rich}, {Simon}, {Sohn}, {Tollerud}, \& {van der Marel}}]{Savino22}
{Savino}, A., {Weisz}, D.~R., {Skillman}, E.~D., {et~al.} 2022, \apj, 938, 101, \dodoi{10.3847/1538-4357/ac91cb}

\bibitem[{{Savino} {et~al.}(2023){Savino}, {Weisz}, {Skillman}, {Dolphin}, {Cole}, {Kallivayalil}, {Wetzel}, {Anderson}, {Besla}, {Boylan-Kolchin}, {Brown}, {Bullock}, {Collins}, {Cooper}, {Deason}, {Dotter}, {Fardal}, {Ferguson}, {Fritz}, {Geha}, {Gilbert}, {Guhathakurta}, {Ibata}, {Irwin}, {Jeon}, {Kirby}, {Lewis}, {Mackey}, {Majewski}, {Martin}, {McConnachie}, {Patel}, {Rich}, {Simon}, {Sohn}, {Tollerud}, \& {van der Marel}}]{Savino23}
---. 2023, \apj, 956, 86, \dodoi{10.3847/1538-4357/acf46f}

\bibitem[{{Schneider} {et~al.}(2001){Schneider}, {Ferrari}, {Matarrese}, \& {Portegies Zwart}}]{schneider01}
{Schneider}, R., {Ferrari}, V., {Matarrese}, S., \& {Portegies Zwart}, S.~F. 2001, \mnras, 324, 797, \dodoi{10.1046/j.1365-8711.2001.04217.x}

\bibitem[{{Shenar} {et~al.}(2024){Shenar}, {Bodensteiner}, {Sana}, {Crowther}, {Lennon}, {Abdul-Masih}, {Almeida}, {Backs}, {Berlanas}, {Bernini-Peron}, {Bestenlehner}, {Bowman}, {Bronner}, {Britavskiy}, {de Koter}, {de Mink}, {Deshmukh}, {Evans}, {Fabry}, {Gieles}, {Gilkis}, {Gonz{\'a}lez-Tor{\`a}}, {Gr{\"a}fener}, {G{\"o}tberg}, {Hawcroft}, {H{\'e}nault-Brunet}, {Herrero}, {Holgado}, {Janssens}, {Johnston}, {Josiek}, {Justham}, {Kalari}, {Katabi}, {Keszthelyi}, {Klencki}, {Kub{\'a}t}, {Kub{\'a}tov{\'a}}, {Langer}, {Lefever}, {Ludwig}, {Mackey}, {Mahy}, {Ma{\'\i}z Apell{\'a}niz}, {Mandel}, {Maravelias}, {Marchant}, {Menon}, {Najarro}, {Oskinova}, {O'Grady}, {Ovadia}, {Patrick}, {Pauli}, {Pawlak}, {Ramachandran}, {Renzo}, {Rocha}, {Sander}, {Sayada}, {Schneider}, {Schootemeijer}, {Sch{\"o}sser}, {Sch{\"u}rmann}, {Sen}, {Shahaf}, {Sim{\'o}n-D{\'\i}az}, {Stoop}, {Toonen}, {Tramper}, {van Loon}, {Valli}, {van Son}, {Vigna-G{\'o}mez}, {Villase{\~n}or}, {Vink}, {Wang}, \& {Willcox}}]{Shenar24}
{Shenar}, T., {Bodensteiner}, J., {Sana}, H., {et~al.} 2024, \aap, 690, A289, \dodoi{10.1051/0004-6361/202451586}

\bibitem[{{Spera} {et~al.}(2015){Spera}, {Mapelli}, \& {Bressan}}]{Spera15}
{Spera}, M., {Mapelli}, M., \& {Bressan}, A. 2015, \mnras, 451, 4086, \dodoi{10.1093/mnras/stv1161}

\bibitem[{{Steidel} {et~al.}(2016){Steidel}, {Strom}, {Pettini}, {Rudie}, {Reddy}, \& {Trainor}}]{steidel16}
{Steidel}, C.~C., {Strom}, A.~L., {Pettini}, M., {et~al.} 2016, \apj, 826, 159, \dodoi{10.3847/0004-637X/826/2/159}

\bibitem[{{Street} {et~al.}(2023){Street}, {Li}, {Khakpash}, {Bellm}, {Girardi}, {Jones}, {Abrams}, {Tsapras}, {Hundertmark}, {Bachelet}, {Gandhi}, {Szkody}, {Clarkson}, {Szab{\'o}}, {Prisinzano}, {Bonito}, {Buckley}, {Marais}, \& {Di Stefano}}]{Street23}
{Street}, R.~A., {Li}, X., {Khakpash}, S., {et~al.} 2023, \apjs, 267, 15, \dodoi{10.3847/1538-4365/acd6f4}

\bibitem[{{Tauris} {et~al.}(2015){Tauris}, {Langer}, \& {Podsiadlowski}}]{Tauris15}
{Tauris}, T.~M., {Langer}, N., \& {Podsiadlowski}, P. 2015, \mnras, 451, 2123, \dodoi{10.1093/mnras/stv990}

\bibitem[{{Tauris} \& {van den Heuvel}(2006)}]{Tauris06}
{Tauris}, T.~M., \& {van den Heuvel}, E.~P.~J. 2006, in Compact stellar X-ray sources, Vol.~39, 623--665, \dodoi{10.48550/arXiv.astro-ph/0303456}

\bibitem[{{Tauris} {et~al.}(2017){Tauris}, {Kramer}, {Freire}, {Wex}, {Janka}, {Langer}, {Podsiadlowski}, {Bozzo}, {Chaty}, {Kruckow}, {van den Heuvel}, {Antoniadis}, {Breton}, \& {Champion}}]{Tauris17}
{Tauris}, T.~M., {Kramer}, M., {Freire}, P.~C.~C., {et~al.} 2017, \apj, 846, 170, \dodoi{10.3847/1538-4357/aa7e89}

\bibitem[{{Telford} {et~al.}(2021){Telford}, {Chisholm}, {McQuinn}, \& {Berg}}]{telford21}
{Telford}, O.~G., {Chisholm}, J., {McQuinn}, K. B.~W., \& {Berg}, D.~A. 2021, \apj, 922, 191, \dodoi{10.3847/1538-4357/ac1ce2}

\bibitem[{{Telford} {et~al.}(2024){Telford}, {Chisholm}, {Sander}, {Ramachandran}, {McQuinn}, \& {Berg}}]{telford24}
{Telford}, O.~G., {Chisholm}, J., {Sander}, A. A.~C., {et~al.} 2024, \apj, 974, 85, \dodoi{10.3847/1538-4357/ad697e}

\bibitem[{{Telford} {et~al.}(2023){Telford}, {McQuinn}, {Chisholm}, \& {Berg}}]{telford23}
{Telford}, O.~G., {McQuinn}, K. B.~W., {Chisholm}, J., \& {Berg}, D.~A. 2023, \apj, 943, 65, \dodoi{10.3847/1538-4357/aca896}

\bibitem[{{Tramper} {et~al.}(2011){Tramper}, {Sana}, {de Koter}, \& {Kaper}}]{tramper11}
{Tramper}, F., {Sana}, H., {de Koter}, A., \& {Kaper}, L. 2011, \apjl, 741, L8, \dodoi{10.1088/2041-8205/741/1/L8}

\bibitem[{{Tramper} {et~al.}(2014){Tramper}, {Sana}, {de Koter}, {Kaper}, \& {Ram{\'\i}rez-Agudelo}}]{Tramper14}
{Tramper}, F., {Sana}, H., {de Koter}, A., {Kaper}, L., \& {Ram{\'\i}rez-Agudelo}, O.~H. 2014, \aap, 572, A36, \dodoi{10.1051/0004-6361/201424312}

\bibitem[{{Urbaneja} {et~al.}(2008){Urbaneja}, {Kudritzki}, {Bresolin}, {Przybilla}, {Gieren}, \& {Pietrzy{\'n}ski}}]{Urbaneja08}
{Urbaneja}, M.~A., {Kudritzki}, R.-P., {Bresolin}, F., {et~al.} 2008, \apj, 684, 118, \dodoi{10.1086/590334}

\bibitem[{{Vanderplas} {et~al.}(2012){Vanderplas}, {Connolly}, {Ivezi{\'c}}, \& {Gray}}]{astroML}
{Vanderplas}, J., {Connolly}, A., {Ivezi{\'c}}, {\v Z}., \& {Gray}, A. 2012, in Conference on Intelligent Data Understanding (CIDU), 47 --54, \dodoi{10.1109/CIDU.2012.6382200}

\bibitem[{{Verbunt}(1993)}]{verbunt93}
{Verbunt}, F. 1993, \araa, 31, 93, \dodoi{10.1146/annurev.aa.31.090193.000521}

\bibitem[{{Vink} {et~al.}(2001){Vink}, {de Koter}, \& {Lamers}}]{Vink01}
{Vink}, J.~S., {de Koter}, A., \& {Lamers}, H.~J.~G.~L.~M. 2001, \aap, 369, 574, \dodoi{10.1051/0004-6361:20010127}

\bibitem[{{Warfield} {et~al.}(2023){Warfield}, {Richstein}, {Kallivayalil}, {Cohen}, {Savino}, {Boyer}, {Garling}, {Gennaro}, {McQuinn}, {Newman}, {Anderson}, {Cole}, {Correnti}, {Dolphin}, {Geha}, {Sandstrom}, {Weisz}, \& {Williams}}]{Warfield23}
{Warfield}, J.~T., {Richstein}, H., {Kallivayalil}, N., {et~al.} 2023, Research Notes of the American Astronomical Society, 7, 23, \dodoi{10.3847/2515-5172/acbb72}

\bibitem[{{Weisz} {et~al.}(2023){Weisz}, {McQuinn}, {Savino}, {Kallivayalil}, {Anderson}, {Boyer}, {Correnti}, {Geha}, {Dolphin}, {Sandstrom}, {Cole}, {Williams}, {Skillman}, {Cohen}, {Newman}, {Beaton}, {Bressan}, {Bolatto}, {Boylan-Kolchin}, {Brooks}, {Bullock}, {Conroy}, {Cooper}, {Dalcanton}, {Dotter}, {Fritz}, {Garling}, {Gennaro}, {Gilbert}, {Girardi}, {Johnson}, {Johnson}, {Kalirai}, {Kirby}, {Lang}, {Marigo}, {Richstein}, {Schlafly}, {Schmidt}, {Tollerud}, {Warfield}, \& {Wetzel}}]{Weisz23}
{Weisz}, D.~R., {McQuinn}, K. B.~W., {Savino}, A., {et~al.} 2023, \apjs, 268, 15, \dodoi{10.3847/1538-4365/acdcfd}

\bibitem[{{Weisz} {et~al.}(2024){Weisz}, {Dolphin}, {Savino}, {McQuinn}, {Newman}, {Williams}, {Kallivayalil}, {Anderson}, {Boyer}, {Correnti}, {Geha}, {Sandstrom}, {Cole}, {Warfield}, {Skillman}, {Cohen}, {Beaton}, {Bressan}, {Bolatto}, {Boylan-Kolchin}, {Brooks}, {Bullock}, {Conroy}, {Cooper}, {Dalcanton}, {Dotter}, {Fritz}, {Garling}, {Gennaro}, {Gilbert}, {Girardi}, {Johnson}, {Johnson}, {Kalirai}, {Kirby}, {Lang}, {Marigo}, {Richstein}, {Schlafly}, {Tollerud}, \& {Wetzel}}]{Weisz24}
{Weisz}, D.~R., {Dolphin}, A.~E., {Savino}, A., {et~al.} 2024, \apjs, 271, 47, \dodoi{10.3847/1538-4365/ad2600}

\bibitem[{{Weisz, Dan}(2024)}]{JWSTSTARS}
{Weisz, Dan}. 2024, JWST Resolved Stellar Populations Early Release Science ("JWSTSTARS"),  STScI/MAST, \dodoi{10.17909/CN6N-XG90}

\bibitem[{{Wellstein} \& {Langer}(1999)}]{wellstein99}
{Wellstein}, S., \& {Langer}, N. 1999, \aap, 350, 148.
\newblock \doarXiv{astro-ph/9904256}

\bibitem[{{Wellstein} {et~al.}(2001){Wellstein}, {Langer}, \& {Braun}}]{Wellstein01}
{Wellstein}, S., {Langer}, N., \& {Braun}, H. 2001, \aap, 369, 939, \dodoi{10.1051/0004-6361:20010151}

\bibitem[{{Willcox} {et~al.}(2023){Willcox}, {MacLeod}, {Mandel}, \& {Hirai}}]{Willcox23}
{Willcox}, R., {MacLeod}, M., {Mandel}, I., \& {Hirai}, R. 2023, \apj, 958, 138, \dodoi{10.3847/1538-4357/acffb1}

\bibitem[{{Williams} {et~al.}(2019){Williams}, {Hillis}, {Blair}, {Long}, {Murphy}, {Dolphin}, {Khan}, \& {Dalcanton}}]{Williams19}
{Williams}, B.~F., {Hillis}, T.~J., {Blair}, W.~P., {et~al.} 2019, \apj, 881, 54, \dodoi{10.3847/1538-4357/ab2190}

\bibitem[{{Williams} {et~al.}(2014){Williams}, {Lang}, {Dalcanton}, {Dolphin}, {Weisz}, {Bell}, {Bianchi}, {Byler}, {Gilbert}, {Girardi}, {Gordon}, {Gregersen}, {Johnson}, {Kalirai}, {Lauer}, {Monachesi}, {Rosenfield}, {Seth}, \& {Skillman}}]{Williams14}
{Williams}, B.~F., {Lang}, D., {Dalcanton}, J.~J., {et~al.} 2014, \apjs, 215, 9, \dodoi{10.1088/0067-0049/215/1/9}

\bibitem[{{Wu} {et~al.}(2023){Wu}, {Qian}, {Li}, {Zejda}, {Mikul{\'a}sek}, {Zhu}, {Liao}, \& {Zhao}}]{Wu23}
{Wu}, C., {Qian}, S., {Li}, F., {et~al.} 2023, \pasj, 75, 358, \dodoi{10.1093/pasj/psad003}

\end{thebibliography}

\end{document}